\documentclass[a4paper,fleqn,usenatbib]{mnras}

\usepackage{newtxtext,newtxmath}
\usepackage[T1]{fontenc}
\usepackage{aecompl}
\usepackage{times} 
\usepackage{graphicx}
\usepackage{rotating}
\usepackage[mathscr]{euscript}
\usepackage{array}
\usepackage[usenames]{color}
\usepackage{epsfig}
\usepackage{amsmath}
\usepackage{natbib}
\def\mbh{\ifmmode{{\mathrm M}_{BH}\,}\else{M$_{BH}$\,}\fi}
\def\msigma{\ifmmode{{\mathrm M}_{BH}-{\sigma}\,}\else{M$_{BH}- \sigma$\,}\fi}
\def\msun{\ifmmode{{\mathrm M}_{\odot}}\else{M$_{\odot}$\,}\fi} 
\def\kms{\ifmmode{{\mathrm{km \, s^{-1}}}}\else{${\mathrm{km \, s^{-1}}}$}\fi} 
\def\kmskpc{km~s$^{-1}$~kpc$^{-1}$}

\def\beq{\begin{equation}}
\def\eeq{\end{equation}}

\def\aj{AJ}%% Astronomical Journal
\def\apj{ApJ}%% Astrophysical Journal
\def\apjl{ApJ}%% Astrophysical Journal, Letters
\def\aap{A\&A}%% Astronomy and Astrophysics
\def\aaps{A\&AS}%% Astronomy and Astrophysics, Supplement
\def\mnras{MNRAS}%% Monthly Notices of the RAS
\def\nat{Nature}%% Nature
\newcommand{\delv}{$\Delta V_{\rm los}$}
\newcommand{\vlos}{$V_{\rm los}$}
\newcommand{\siglos}{$\sigma_{\rm los}$}

\usepackage{graphicx}	% Including figure files
\usepackage{amsmath}	% Advanced maths commands
\usepackage{amssymb}	% Extra maths symbols

\title[X-shape in the Milky Way Bar]{On the orbits that generate the X-shape in the Milky Way bulge}

\author[Abbott et al.]{Caleb G. Abbott,$^{1,2}$\thanks{E-mail: calebga@umich.edu}
Monica Valluri,$^{1}$\thanks{E-mail:mvalluri@umich.edu}
Juntai Shen$^{3,4}$\thanks{E-mail:jshen@shao.ac.cn}
and Victor P. Debattista$^{5}$\thanks{E-mail: vpdebattista@gmail.com}
\\
$^{1}$Department of Astronomy, University of Michigan, Ann Arbor, MI 48109, USA\\
$^{2}$Department of Physics and Astronomy, Georgia State University, Atlanta, GA 30303, USA\\
$^{3}$Key Laboratory for Research in Galaxies and Cosmology, Shanghai Astronomical Observatory, Chinese Academy of Sciences, Shanghai 200030, China\\
$^{4}$College of Astronomy and Space Sciences, University of Chinese Academy of Sciences, 19A Yuquan Road, Beijing 100049, China\\
$^{5}$Jeremiah Horrocks Institute, University of Central Lancashire,  Preston, PR1 2HE, UK
}

\date{Accepted XXX. Received YYY; in original form ZZZ}

\pubyear{2016}

\begin{document}
\label{firstpage}
\pagerange{\pageref{firstpage}--\pageref{lastpage}}
\maketitle

\begin{abstract}
The Milky Way bulge  shows a box/peanut or  X-shaped bulge (hereafter BP/X) when viewed in infrared or microwave bands. We examine orbits in an N-body model of a barred disk galaxy that is scaled to match the kinematics of the Milky Way (MW) bulge. We generate maps of projected stellar surface density, unsharp masked images, 3D excess-mass distributions (showing mass outside ellipsoids), line-of-sight number count distributions, and 2D line-of-sight kinematics for the simulation as well as co-added orbit families, in order to identify the orbits primarily responsible for the BP/X shape.  We estimate that between 19-23\% of the mass of the bar  {in this model} is associated with the BP/X shape and that the majority of bar orbits contribute to this shape which is clearly seen in projected surface density maps and 3D excess mass for non-resonant box orbits, ``banana'' orbits, ``fish/pretzel'' orbits  and  ``brezel'' orbits. Although only the latter two families (comprising 7.5\% of the total mass) show a distinct X-shape in unsharp masked images, we find that nearly all bar orbit families contribute some mass to the 3D BP/X-shape.  All co-added orbit families show a bifurcation in stellar number count distribution with distance that resembles the bifurcation observed in red clump stars in the MW. However, only the box orbit family shows an increasing separation of peaks with increasing galactic latitude $|b|$, similar to that observed. Our analysis  {suggests} that no single orbit family fully explains all the observed features associated with the MW's BP/X shaped bulge, but collectively the non-resonant boxes and various resonant boxlet orbits contribute at different distances from the center to produce this feature.  We propose that since box orbits (which are the dominant population in bars) have three incommensurable orbital fundamental frequencies, their 3-dimensional shapes are highly flexible and, like Lissajous figures, this family of orbits is most easily able to adapt to evolution in the shape of the underlying potential. 
\end{abstract}


%%%%%%%%%%%%%%%%%%%%%%%%%%%%%%%%%%%%%%%%%%%%%%%%%%

%%%%%%%%%%%%%%%%% BODY OF PAPER %%%%%%%%%%%%%%%%%%

\section{Introduction}

Observations of bars in edge-on extragalactic  disk galaxies often show a distinct boxy-peanut (BP) shaped bulge, which reveals a clear X-shaped structure when the images are subjected to unsharp masking.  These structures are now observed in about 45\% of edge-on disk galaxies \citep{burbidge_59,shaw_87,com_etal_90,lutticke_etal_00,lauri_etal_11}, which if one accounts for the range of possible viewing angles, suggests that these structures are very common.

In the Milky Way (MW), \citet{blitz_spergel_91} inferred the existence of a bar from 2.4micron observations of the Galactic Center region. A peanut shaped bulge was first clearly seen in the multi-parameter model of the COBE/DIRBE images of the Galactic Bulge \citep{freundenreich_98}. Further evidence for an X-shaped bulge was inferred from the bifurcation in red clump (RC) star counts in the 2MASS and OGLE-III surveys \citep{skrutskie_06,mcwilliam_zoccali_10,nataf_10} and confirmed by \citet{ness_etal_12}.  These observations showed that at the Galactic coordinates $l = 0^\circ$ and $|b| > 5^\circ$ the distribution of the RC stars splits into two distinct peaks, a bright peak on the near side of the Galactic center (GC), referred to as the ``bright red clump'' (BRC) and fainter peak on the far side of the GC, the ``faint red clump" (FRC).  \citet{mcwilliam_zoccali_10} interpreted these two clumps as evidence for an X-shape within the MW bar.  This was confirmed shortly thereafter by \citet{saito_etal_11}. An excellent demonstration of the existence of a X-shaped bulge is seen in the 3D distribution of  red clump stars from the VVV survey \citep{wegg_gerhard_13}.  Recently \citet{ness_lang_16} have shown, by constructing an image of the Milky Way bulge from an independent co-adding of publicly available WISE data, that Milky Way bulge shows a distinct X-shaped structure in the projected stellar distribution even without unsharp masking.

 Using Fourier harmonic fitting to the isophotal distributions of the sample of BP/X galaxies  \citet{ciambur_graham_16} recently showed that the peanut/X-shapes embedded in near edge-on disks are best described not by the 4th Fourier harmonic ($B_4$), which is usually used to distinguish between boxy and disky bulges, but by the 6th Fourier harmonic ($B_6$). Furthermore they use five quantitative metrics to describe the strength of the BP/X shape and show that the length and strength of the peanut increases with the rotation velocity of the disk. 

 BP bulges with vertical X-shaped structures are found to arise naturally in simulations of bars as a result of the asymmetric buckling instability \citep{Raha91,oneill_dubinski_03,Martinez-V_Shlosman_04,debattista_etal_05,bureau_athanassoula_05,Debattista06}. This instability typically occurs as the bar strengthens and when the velocity dispersion along the length of the bar ($\sigma_x$) significantly exceeds the vertical velocity dispersion ($\sigma_z$),  \citep[e.g. when $\sigma_z/\sigma_x \lesssim 0.4$,][]{araki_phd,Martinez-V_etal06}.  The instability causes a redistribution of kinetic energy from the plane of the disk resulting in a significant increase of the thickness of the bar, giving it  the familiar BP shape \citep{pfenniger_friedli_91}.  The X-shaped structure seen in projection and in unsharp masked images is probably more peanut-like in 3D, and the visual perception of an X-shape in projection is enhanced by the pinched, concave inner isodensity contours \citep{li_shen_15}.  This BP/X-shape can qualitatively reproduce  the observed bimodal distributions  in the number counts of RC stars along the line of sight that were used as evidence for the discovery of the X-shape. Indeed, several studies comparing N-body simulations of bars with the observed spatial  distributions and line-of-sight velocity distributions of RC stars in the MW \citep{shen_etal_10, li_shen_12, vasquez_etal_13, gardner_etal_14, nataf_etal_15, qin_etal_15} argue that the buckling of bars which gives rise to the BP shape seen in external galaxies is also responsible for the observed X-shaped structure in the MW bulge. Alternative mechanisms such as resonant trapping of stars on vertical inner Lindblad resonances  \citep{com_san_81,com_etal_90,quillen02,quillen_etal_14} and resonant trapping of disk material around the stable 3D periodic orbits associated with vertical 2:1 and 4:1 resonances  \citep{patsis_xilouris_06} may also contribute. While most boxy bulges are generally associated with bars, axisymmetric boxy bulges can also exist \citep{rowley_88,patsis_etal_02nonbarred}.

An obvious question that arises is whether one or more specific orbit families  in self-consistent bars are responsible for the BP bulge and/or the X-shape.  The standard view of orbital structure of bars has held that the dominant families of bar orbits are quasi-periodic or regular orbits that arise from stable periodic prograde x1 and x2 orbits \citep[e.g.][]{contopoulos_papayannopoulos_80,athanassoula_etal_83}. For example the vertical bifurcations of the x1 family, e.g. the 2:-2:1\footnote{We use orbital fundamental frequencies in Cartesian coordinates to define resonant orbits: i.e. orbits that are resonant satisfy  the resonant condition $\Omega_x:\Omega_y:\Omega_z= l:m:n$ where $l,m,n$ are small integers. The sign  associated with an integer depends on the signs of the slope and intercept of a resonance line when plotted on a frequency map.} resonant orbit family   {(referred to hereafter as the ``banana'' orbit family, often referred to in the literature as ``x1v1'' orbits )} is widely considered to be  the backbone of BP/X bulges seen in edge-on buckled bars \citep{pfenniger_friedli_91,skokos1,patsis1,Athanassoula05}.  \citet{skokos2} also found that in a model without 2:-2:1 orbits, the bar was supported by orbit families connected with $z$-axis orbits.  \citet{Patsis_Katsanikas_14a, Patsis_Katsanikas_14b} also present a dynamical mechanism for building X-shaped peanuts with families of periodic orbits that are not bifurcations of x1 orbits. 

Recently \citep[e.g.][]{qin_etal_15} it has been argued that it is difficult to clearly identify a single orbit family that reproduces both the spatial and kinematical distributions associated with the X-shape in the MW bulge.
\citet{portail_etal_15a} constructed  models of the Galactic bulge that fit the spatial distribution of bulge stars from the VVV survey \citep{wegg_gerhard_13} and kinematical data from the Bulge RAdial Velocity Assay  \citep[BRAVA][]{rich_etal_07,kunder_etal_12}. By analyzing the orbital structure of their self-consistent Made-to-Measure (M2M) models and N-body models,  \citet{portail_etal_15b} argued that since the 2:-2:1 vertical resonance of the x1 orbit family appears primarily  in the outer parts of  bars, they cannot explain the X-shape in the Milky Way bar which is observed over a range of distances from the center of the bar. Instead they proposed that  resonant boxlet orbits associated with the 3:0:-5 resonance (which they term ``brezel'' orbits) are primarily responsible for the X-shape. They further conclude that  $\sim$40-45\% of the stellar orbits in the bulge/bar contribute to the X-shape.

The fraction of stellar mass  associated with the X-shape has been estimated from unsharp masking by \citet{li_shen_12} to about 7\% of the bulge mass. In contrast \citet{portail_etal_15a}  estimate the 'excess mass' lying outside ellipsoids and infer that 20-25\% of the mass of the bulge is associated with the X-shape.  These differences are most probably due to the differences in the methods used to compute the mass associated with the X-shape.

We recently carried out a comprehensive analysis of a representative sample of orbits drawn from the self-consistent particle distribution  of two N-body bars \citep[][hereafter V16]{valluri_etal_16}.  We showed that the dominant bar orbit family (comprising $\sim$ 60\% of bar orbits) is the box orbit family (otherwise referred to as the ``non-resonant x1 orbit family'' in 3D potentials) which originates from perturbations of the linear long-axis orbit. This box orbit family is well known from studies of stationary triaxial ellipsoids where it is the dominant family \citep{dezeeuw_85b,BT08}.  In the frame co-rotating with an N-body bar, these box orbits are modified only slightly by the pseudo-forces arising from the rotating potential. Furthermore, V16 showed that the  {vertical bifurcation of the x1 orbit, the 2:-2:1 ``banana'' orbits comprises only about 3\% of all bar orbits} and is found primarily in the outer half of the bar. This latter result is consistent with the findings of \citet{portail_etal_15b}. V16 found that the most important resonant boxlet family (comprising about $\sim $6\% of orbits) in  N-body bars\footnote{V16 showed that this is also the most important resonant boxlet family in rapidly rotating prolate triaxial potentials described by the Dehnen profile \citep{dehnen_93}.} is associated with the 3:-2:0 resonance that we refer to as ``fish/pretzel'' orbits. N-body bars were also found to contain long-axis tube orbits (4-8.5\%), a small fraction of short-axis tube orbits (i.e. orbits originating from retrograde x4 orbits and prograde x2 orbits) and a significant fraction (18-22\%) of chaotic orbits. The precise fractions of orbits associated with different families are expected to be  somewhat dependent on the details of the bar potential, with the range of orbital fractions listed above arising in the two models discussed in V16.
  
In this paper we re-examine  regular orbits in one of the N-body simulations previously classified and analyzed in V16. Although we found a significant fraction of chaotic orbits in our N-body bars we do not examine them here because their spatial distributions evolve with time implying that the inferred spatial distribution is sensitive to the choice of orbital integration time. However, we note in passing that on short  orbit integration times the majority of chaotic orbits in our models behave like non-resonant box orbits. 

We use four different diagnostics to compare the full simulation with co-added orbits from different orbit families: (1) projected surface density distribution and unsharp masked images of these density distributions; (2)   the 3-dimensional  ``excess-mass'' that lies outside concentric ellipsoids;  (3) bifurcation in the stellar number count distributions along several lines-of-sight and variation of the separation between the peaks of these distributions as a function of galactic latitude $b$; (4)  2D maps of the separation in the mean line-of-sight velocity (\delv) for red clump stars on the near and far sides of the GC. The goal of this analysis is to determine the types of orbits that give rise to the box/peanut and/or the X-shapes observed in both N-body bars and real galactic bars. Although our models have not been specifically tailored to fit the Milky Way they describe several of the observed features extremely well,  and therefore yield useful insights into observations of structural and kinematical features in the Milky Way bulge. 

This paper is organized as follows. In Section~\ref{sec:methods} we describe the  bar simulations from which the line-of-sight kinematics were derived and in which orbits were computed. We also briefly recap the main results of V16 describing the orbit families in  N-body bars. In Section~\ref{sec:results} we present comparisons between the full simulation and co-added orbits in different families and qualitatively compare these with Milky Way data where available. We summarize our results and conclude in Section~\ref{sec:concl}.

\section{Simulations and  analysis methods}
\label{sec:methods}

\subsection{N-body Bar Models}
\label{sec:models}

%%%% Figure 1 -- Chisquare of it to BRAVA kinematics (model A, C) -- %%%%%%
\begin{figure}
\centering
	\includegraphics[width=.36\textwidth]{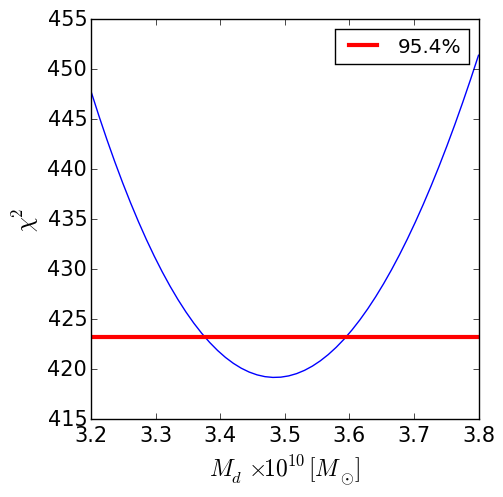}
	\includegraphics[width=.38\textwidth]{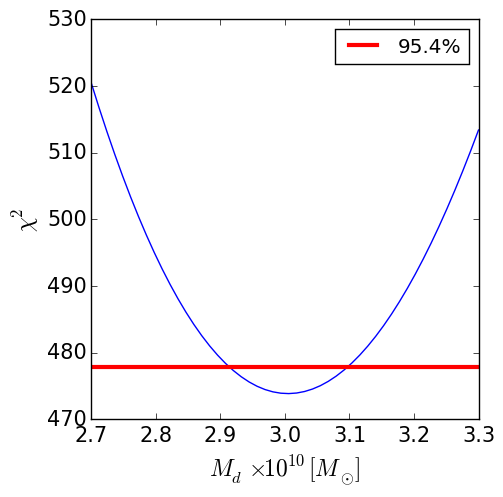}
\caption{$\chi^2$ of the fit to the BRAVA kinematical data \citep{rich_etal_07,kunder_etal_12} shown in Fig.~\ref{fig:Bra_fitAB} as a function of $M_d$ for Model A  (top) and Model C (bottom). Horizontal red lines  show the $3\sigma$ confidence intervals in $\Delta\chi^2$ (relative to the minimum) for one degree of freedom.  The minimum $\chi^2$  was obtained for $M_d= 3.48\times10^{10}$ \msun (for Model A) and for $M_d= 3.01\times10^{10}$ \msun (for Model C).}
    \label{fig:chisqBra_fitAB}
\end{figure}
%%%% Figure 1 -- Chisquare of it to BRAVA kinematics (model A, C) -- %%%%%%

%%%%%% Figure 2 -- Fit to BRAVA kinematics (model A, C) -- %%%%%%

\begin{figure*}
	\includegraphics[width=.30\textwidth]{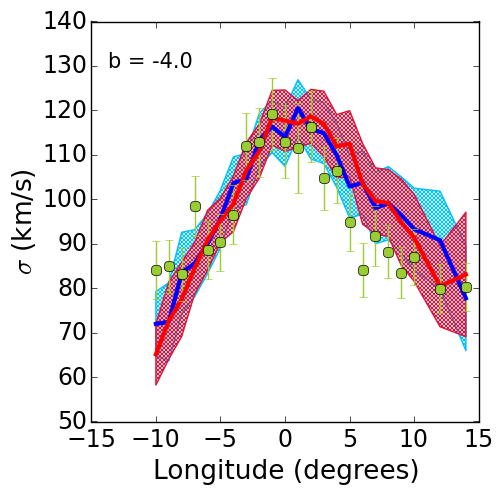}
	\includegraphics[width=.30\textwidth]{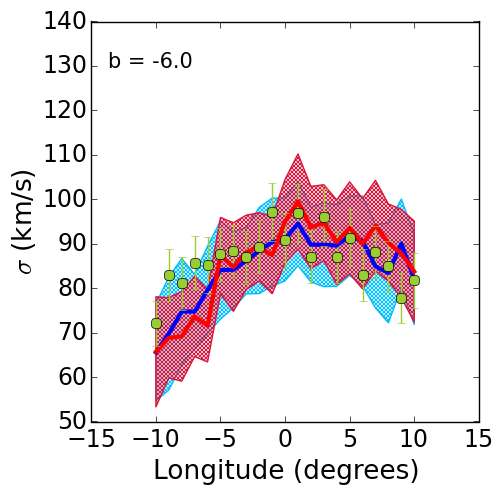}
	\includegraphics[width=.30\textwidth]{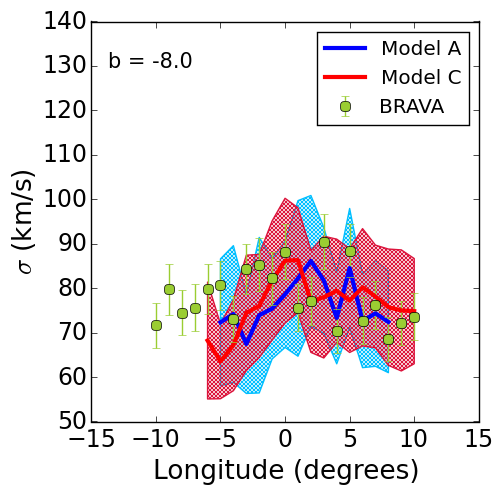}
	\includegraphics[width=.30\textwidth]{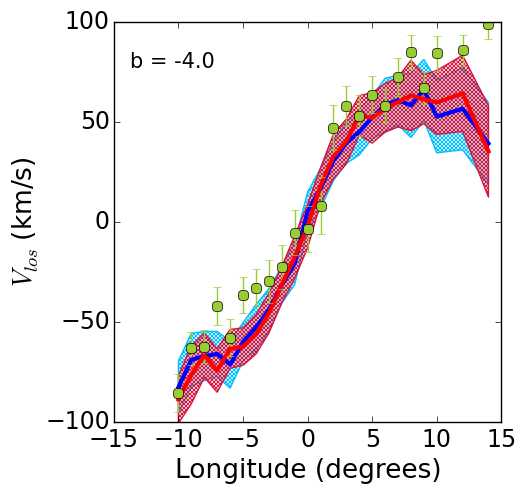}
	\includegraphics[width=.30\textwidth]{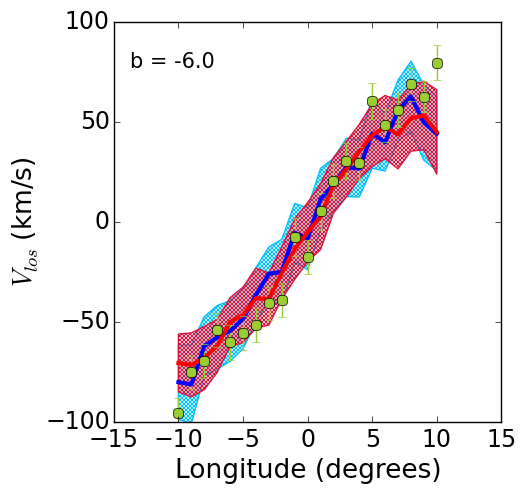}
	\includegraphics[width=.30\textwidth]{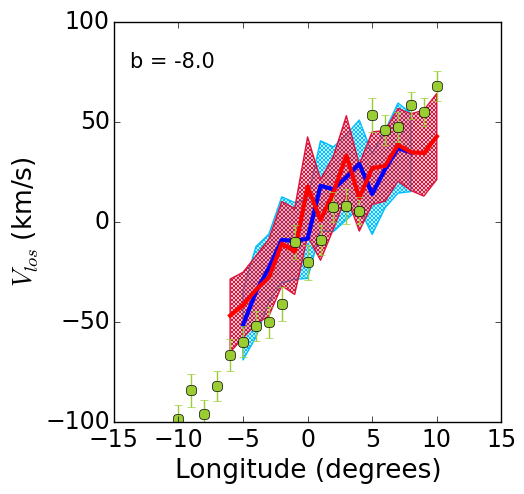}
\caption{The fit to BRAVA data (green points) for the best-fit $M_d$ values are shown for $b=-4^\circ, b=-6^\circ, b = -8^\circ$ for Model A (blue solid curves with light blue 3$\sigma$ bootstrap error bands) and Model C (red solid curves with pink 3$\sigma$ bootstrap error bands).  Bins with fewer than 100 particles (e.g. for $b = -8^\circ$, $ -12^\circ< l < -6^\circ$) were not included in the fit.}
    \label{fig:Bra_fitAB}
\end{figure*}
%%%%%% Figure 2 -- Fit to BRAVA kinematics (model A, C) -- %%%%%%

We examine two  N-body bar models both arising from the same set of initial conditions. The models are almost identical to those used by one or more authors of this paper in the past \citep{shen_sellwood_04,brown_etal_13,valluri_etal_16}.  What follows is a brief description of the simulations and the orbit analysis methods employed. For a more detailed description of the simulation methods and the initial conditions the reader is referred to the above papers.  The initial disk galaxy contains $\sim 2.8$~million particles described by a Kuzmin density profile with disk mass $M_d$  and radial length scale $R_d$.  Particles were given initial velocities such that the Toomre parameter is $Q \sim 1.5$, making the disk unstable to bar formation. The disk is embedded in a static, spherical dark matter halo described by a logarithmic potential. While it is well known that there are differences in the growth rate and strength of bars that form in live dark matter halos and static dark matter halos \citep{Athanassoula02}, \citet{shen_sellwood_04} found little difference in the evolution of bar orbits in live and static dark matter halos with the same density profile. Furthermore  a recent  study of a bar is a live halo with a cusp finds substantially similar orbital structure \citep{gajda_etal_16} to that found by V16, hence our results are expected to be fairly typical for bars of similar strength to those studied here.

The initial conditions were evolved using a three-dimensional, cylindrical, polar grid-based N-body code \citep{sellwood_valluri_97,sellwood14}. The bar forms,  grows in strength and then buckles. After the buckling phase the bar strength saturates at $t \sim 700$ time units and thereafter the bar maintains a nearly steady rotation speed and bar strength. A frozen snapshot of the bar at $t=700$ units is therefore used to represent a pure bar and is referred to as ``Model A''. This simulation is evolved further as a central point mass (representing a supermassive black hole with a $M_{\mathrm{BH}}$ = .0002 $M_d$) is grown adiabatically  at the center. Transients associated with the growth of the central point mass dissipate by $t=1200$. The snapshot of the simulation at this time is referred to as ``Model C''.%\footnote{V16 also examined orbits in a simulation with a ten times more massive central point mass referred to as ``Model B''.}

Following standard practice the coordinate system for the model is defined with the $x$-axis along the length of the bar, the $y$-axis is in the plane of the disk perpendicular to the $x$-axis, and the $z$-axis is perpendicular to the disk. 

The simulations and co-added orbits were analyzed for two different orientations of the bar: (a) an edge-on disk very far away from the observer with the bar oriented perpendicular to the line-of-sight to the galaxy (i.e. bar seen side-on), and (b)  a configuration designed to mimic the orientation of the MW bar as seen from the Sun (referred to hereafter as ``the heliocentric rotated" (HCR) frame).  

In the HCR frame the simulated bar was rotated such that the long-axis of the bar lies at 27$^\circ$ to our line-of-sight to the GC to match the orientation of the MW bar  of $27^\circ\pm2^\circ$\citep{wegg_gerhard_13}. The distance of the Sun from the GC is assumed to be  8~kpc  \citep[e.g.][]{Eisenhauer_etal_03}.  In this frame  the bar is not perpendicular to our line-of-sight to the GC, hence we defined a new coordinate system:  ($\beta$, $\alpha$, $z$) with the origin at the GC and with the positive $\alpha$-axis oriented along the line connecting the GC  to the Sun and the $\beta$-axis perpendicular to our line of sight to the GC with positive $\beta$ values associated with positive longitudes. The definition of the $z$-axis is the same in both coordinate systems.

%%%%%% Figure 3 -- Model A,C, 2D Projected density, unsharp masked -- %%%%%%

\begin{figure*}
	\includegraphics[trim=0.pt 0.pt 0.pt 0.pt ,clip,width=.7\textwidth]{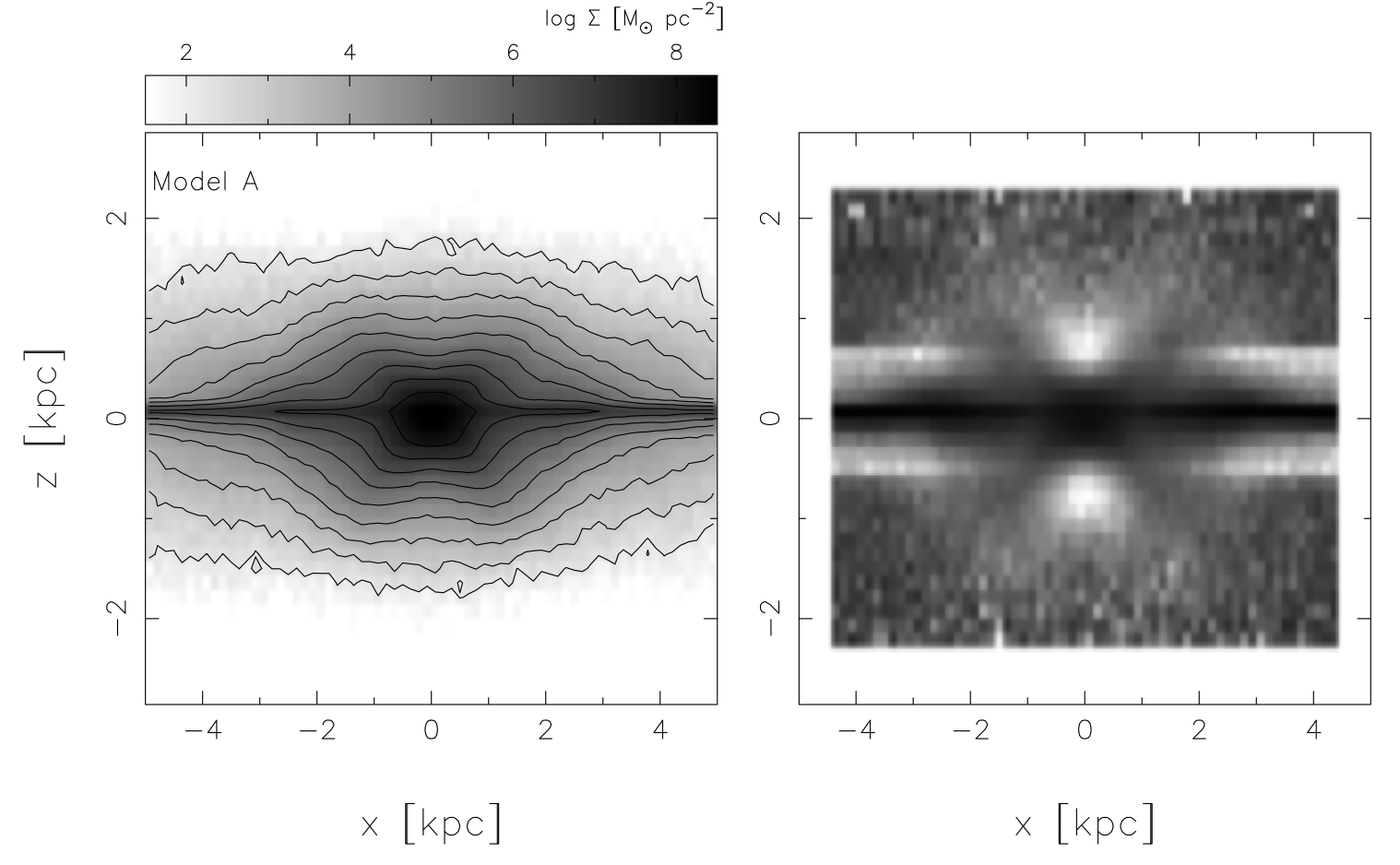}
        \includegraphics[trim=0.pt 0.pt 0.pt 0.pt ,clip,width=.7\textwidth]{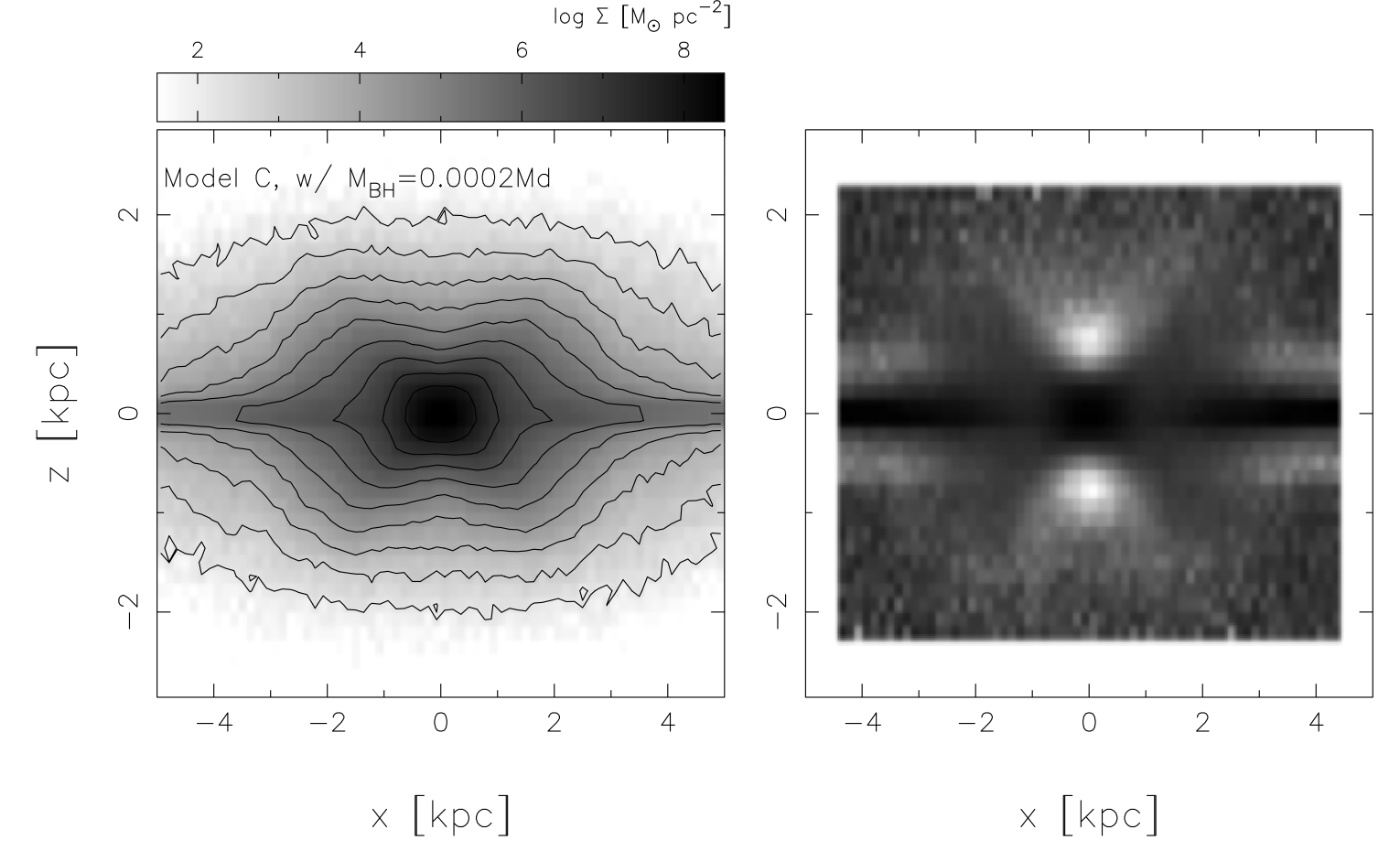}
   \caption{Projected surface mass density (left column) and unsharp masked images (right column) to highlight the X-shape in Model A (top row) and Model C (bottom row) when the bar is observed  in an edge-on disk with bar perpendicular to our line-of-sight.
    \label{fig:unsharpfullsims}}
\end{figure*}
%%%%%%%%%% 2d Projection and unsharp mask

The simulations were run in units with $G = M_d = R_d = 1$ where $G$ is Newton's gravitational constant.  Using standard dimensional analysis, the unit of time is $t_{\mathrm{dyn}} =(R_d^3/GM_d)^{1/2}$.  In order to compare the simulations with observations of the MW bar/bulge it was necessary to convert the simulation units to physical units.   The bar length was estimated using two different methods. The first method measures the $m=2$ Fourier moment amplitude and the second method measures the  phase in annuli. A lower limit on the length of the bar is obtained by determining the length at which the strength of the $m=2$ mode drops to  20\% of the peak amplitude.  While this is a reasonable estimate it is generally considered to be a lower limit to the bar length.  The measurement based on the phase determines the bar length to be the radius at which the phase deviates by more than $10^\circ$ from a constant.  This tends to overestimate the bar length. We use a simple average of both estimates to set the length of the bar to 3.9 units in  Model A, and 4.5 units in Model C. We compare these values to recent estimates of the length of the MW bar \citep[$5\pm0.2$~kpc for the full bar,][]{wegg_gerhard_portail_15}, and thereby choose $R_d$  to be 1.28~kpc for Model A and 1.1~kpc for Model C.

%%%%%% Figure 4a -- Model A,C, 2D projection of excess mass -- %%%%%%
\begin{figure*}
	\includegraphics[trim=0.pt 0.pt 0.pt 0.pt ,clip,width=.3\textwidth]{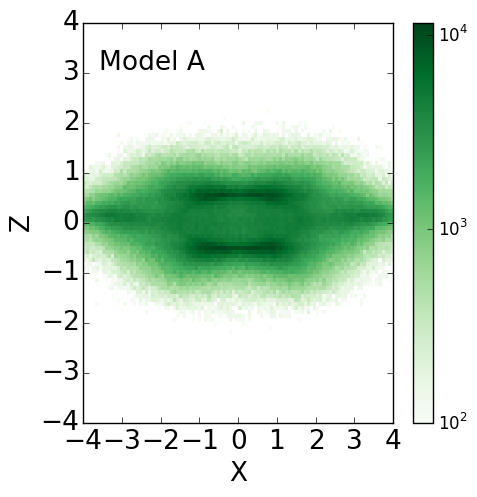}
        \includegraphics[trim=0.pt 0.pt 0.pt 0.pt ,clip,width=.3\textwidth]{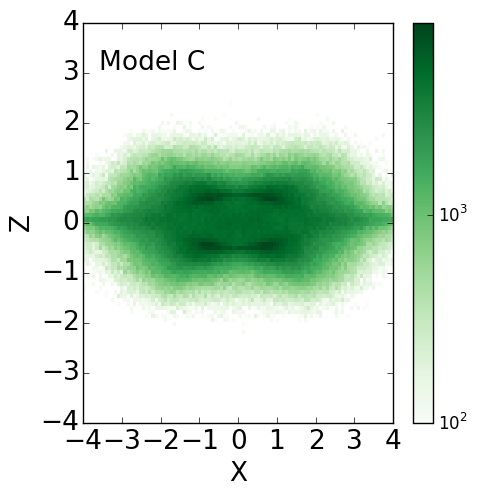}
   \caption{{2D projection of the 3D excess mass density distributions for Model A (left) and Model C (right). The excess mass outside ellipsoids (see text for details) shows a clear BP/X shape in both models.  The color gradient as shown in the bar on the right of each plot is in units of $\msun\ / kpc^{2}$.}
    \label{fig:model_Xmass}}
\end{figure*}

%%%%%% Figure 4a -- Model A,C, 2D projection of excess mass -- %%%%%%

%%%%%% Figure 5-- Model A,C, Projected density, hist, CDF, 2D kin -- %%%%%%
\begin{figure*}
\centering
	\includegraphics[width=.235\textwidth]{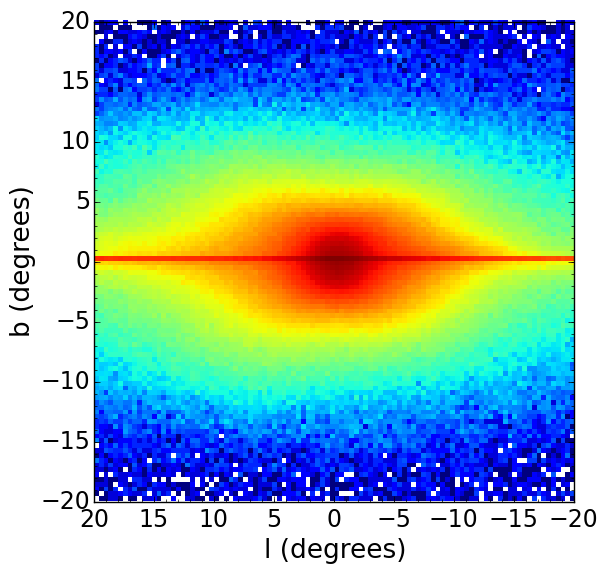}
	\includegraphics[width=.22\textwidth]{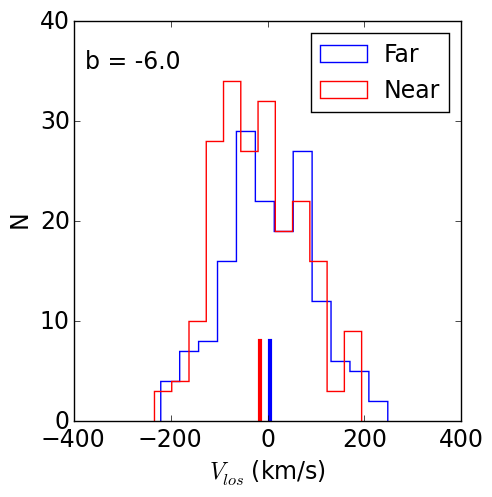}
	\includegraphics[width=.21\textwidth]{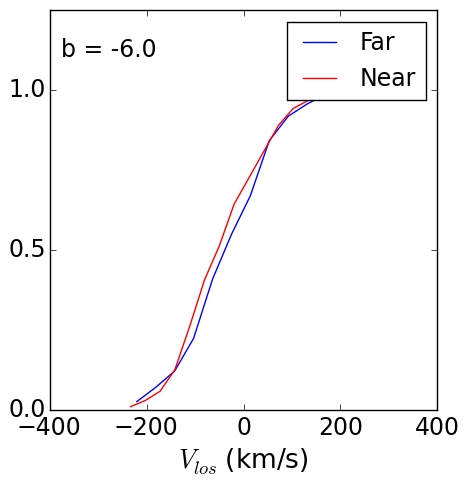}
	\includegraphics[width=.32\textwidth]{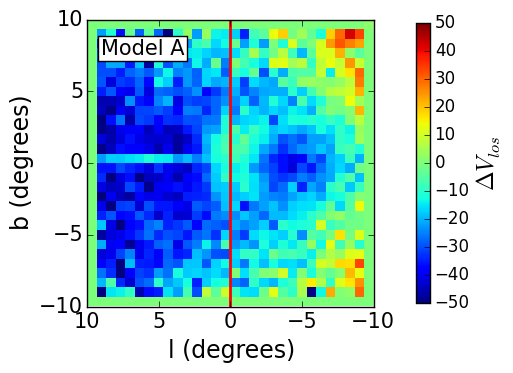}
	%\hspace{2cm}
	%\includegraphics[width=.20\textwidth]{mean_-4b_0l_7_01_hist.png}
	%\includegraphics[width=.20\textwidth]{mean_-6b_0l_7_01_hist.png}
%	\includegraphics[width=.24\textwidth]{12_05_unrot_dist.png}
	\includegraphics[width=.235\textwidth]{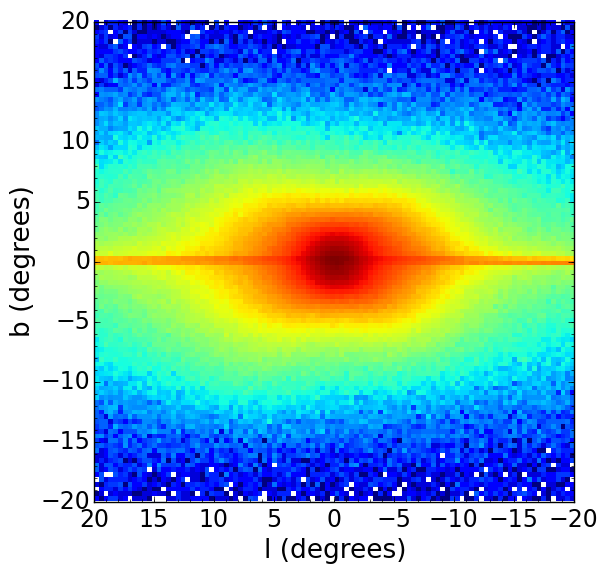}
	\includegraphics[width=.22\textwidth]{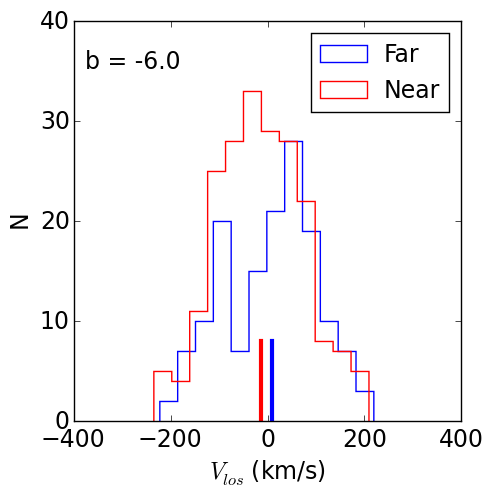}
	\includegraphics[width=.21\textwidth]{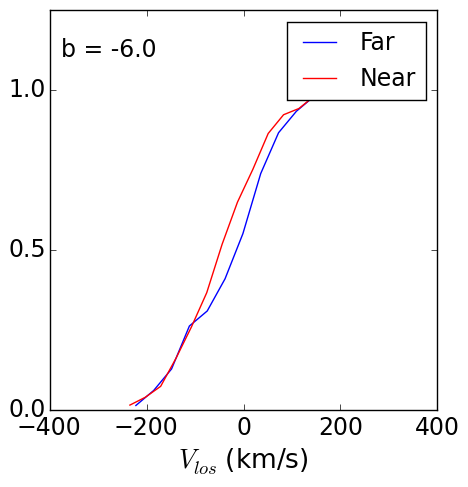}
	\includegraphics[width=.32\textwidth]{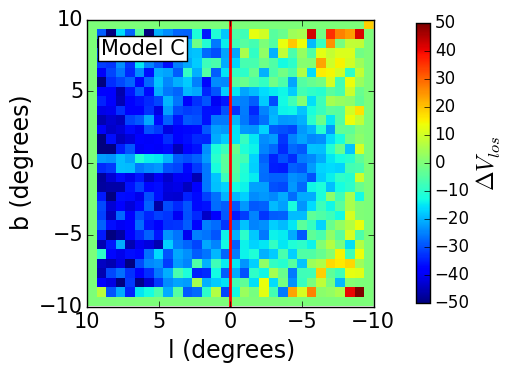}
	\hspace{2cm}	
    \caption{The first column shows projected stellar distribution of the two bars (Model A: top row, Model C: bottom row) oriented at 27$^\circ$ to our line-of-sight to the GC. An asymmetry similar to that observed for star counts in the Milky Way bar \citep[e.g.][]{wegg_gerhard_portail_15} is clearly seen. The second column shows histograms of \vlos\, of stars at $(l,b) = (0^\circ,-6^\circ)$ for stars on the `Near' and `Far' sides of the GC shown in  red and blue histograms respectively. The mean velocities of these distributions (computed via Gaussian fitting) are marked by short red and blue vertical lines. The third column shows the CDF for these velocity histograms.  The red CDFs rise slightly more rapidly than the blue CDFs showing that the distribution of stars on the near side of the GC peaks at more negative \vlos\, than stars on the far side.   The final column shows two-dimensional maps of \delv  (see text for definition) which show the characteristic asymmetric ring like structure with negative \delv values surrounding a region with \delv$\sim 0$. }
    \label{fig:Kin_A}
\end{figure*}
%%%%%% Figure 5 -- Model A,C, Projected density, hist, CDF, 2D kin -- %%%%%%

  The parameter representing the disk mass, $M_d$ was determined by fitting the simulations to the MW bulge line-of-sight kinematical data (\vlos\, and velocity dispersion \siglos) obtained by the BRAVA survey at 77 different pointings \citep{rich_etal_07,kunder_etal_12}.  We determine the best fit value of $M_d$ by varying this quantity (with $R_d$ fixed as described above) and computing the resulting $\chi^2$ of the fit to all the kinematical data \citep[in a manner similar to that used by][hereafter G14]{gardner_etal_14}. This was done separately for Model A and Model C since the growth of the black hole and the angular momentum transfer induced by the bar, increases the density of stars (therefore the velocity dispersion) in the nuclear region. Figure~\ref{fig:chisqBra_fitAB} shows  $\chi^2$ of the fit to all the kinematical data as a function of $M_d$ for Model A (top) and Model C (bottom). The horizontal red lines mark the 3-$\sigma$ $\Delta \chi^2$ error region. The best fit values of $M_d$ were $3.48\times10^{10}$ ($\pm 5\times10^{8}$) \msun\, (Model A), and  $3.01\times10^{10}$ ($\pm 5\times10^{8}$)\msun\, (Model C).  In Model C this value of $M_d$ gives a $M_{\mathrm{BH}} = 6.02\times10^{6}$ \msun\, making it roughly 1.6 times the mass of the MW black hole.  With these values for $R_d$ and $M_d$, $t_{\mathrm{dyn}} = 3.69$~Myr and $3.17$~Myr for Model A and C respectively. The $\chi^2$ of the fit to kinematics for the two models are $\chi^2 = 419$ (Model A) and $\chi^2=473$ (Model C) giving reduced $\chi^2$ of 2.7 and  3.1 respectively signifying  only moderately good  fits to the data. We note that since the bar simulations analyzed in this paper were not explicitly designed to fit the Milky Way bar,  the orbit distributions in our simulations may not exactly represent those in the Milky Way.
 
 Figure~\ref{fig:Bra_fitAB} shows the BRAVA data (\vlos\, and \siglos, green points) at $l=0^\circ$ and $b= -4^\circ, -6^\circ, -8^\circ$, with the best fitting models shown as solid curves with 3$\sigma$ bootstrap error bands (Model A blue/light blue,  Model C red/pink). Bootstrap errors on the fits were obtained by taking 100 random re-samplings of simulation particles in each bin (allowing for repetition)  that contained at least 100 simulation particles. Bins with fewer than 100 particles were not fitted.

Figure~\ref{fig:unsharpfullsims} shows  the projected surface density distributions and unsharp masked images (which highlight the X-shape). Unsharp masked images were obtained by taking the full projected particle distribution, binning the particles on a 70$\times$70 pixel grid, and then smoothing the binned density distribution with a square kernel (4 pixels wide) that is moved over the entire grid. The smoothed image in then subtracted from the original image to obtain the unsharp masked image. The unsharp masked images are re-scaled to provide the best image contrast.

Both models shows a clear BP/X shaped bulge in projected surface density (left hand column) and the unsharp masked images (right hand column) clearly reveal an off-center X-shape \citep[according to the classification of][]{bureau_etal_06}. We note that although the off-center X-shape is also observed in the Milky Way, the strength of the X-shape in our bars maybe somewhat weaker than observed in the Milky Way \citep{wegg_gerhard_13,portail_etal_15a}, hence the detailed orbital structure of the MW bar may differ somewhat from the one presented here.

Following a method similar to that discussed in \citet{portail_etal_15a} we construct 3-dimensional isodensity surfaces for both of our N-body models. This is done by binning all the particles in the bar on a 3D Cartesian grid comprising of 40$\times$40$\times$40 bins.   We then identify 10 equally spaced isodensity levels. Starting from the highest isodensity surface and working outwards for each surface, we define the largest ellipsoid that can be enclosed entirely within a given isodensity surface. The mass enclosed within this ellipsoid is subtracted from the mass enclosed within the isodensity surface to obtain the residual ``excess mass'' associated with the BP/X shape. This mass was found to be 23\% of the total mass of the bar for Model A and 19\% for Model C, consistent with values found by \citet{portail_etal_15a} ($24^{+5}_{-4}$\%).  This excess mass distribution projected on to the $x-z$ plane is shown in Figure~\ref{fig:model_Xmass}.  We find that the residual mass distribution in both Models A and C is peanut shaped unlike the residual mass distribution shown in \citet[see Figure~18 in][]{portail_etal_15a} which shows a distinct X-shape in the $x-z$ projection. However our results are completely consistent with the results obtained by  \citet{li_shen_15}.

The terms ``box/peanut'' and ``X-shaped bulge'' are generally used interchangeably in the literature. Our analysis above shows that while this is probably justified, the process by which this structure is identified, e.g. direct examination of projected density, unsharp masking or analysis of 3D spatial distribution,  appears to affect conclusions about how much mass is associated with it.

\citet{vasquez_etal_13}  obtained line-of-sight velocities and proper-motions for stars belonging to the BRC and the FRC.  These authors found that the two  arms of the X-shape intersect the line-of-sight from the Sun to the GC at $l = 0^\circ$ and $|b| > 4^\circ$.  %This overlap in the arms of the X-shape, along with the distance between each arm, causes the counts of red clump (RC) stars from 2MASS (Skruskie et al. 2006 \ca{look up this paper Caleb)} and OGLE-III (Szymanski et al. 2011 \ca{again, look this up}) to show a split in the RC into two components; implying two peaks in the stellar density at (l,b) examined. 
\citet{vasquez_etal_13} found that the velocity distribution of stars in the BRC is skewed towards negative \vlos\ while the velocity distribution of stars in the FRC is skewed towards positive \vlos. Cumulative distribution functions (CDF) of the heliocentric radial velocities of the BRC and FRC stars show that the median velocities of these two distributions differ by $\sim 50$\kms. Their comparison with a simulated bar \citep[from][]{debattista_etal_05} (see also G14) shows that this difference in the CDFs of `Near' and `Far' side stars is a characteristic of the X-shape.

We searched for a similar feature in \vlos\ in our simulation by splitting stars in our simulations into a near group and a far  group  by defining star particles as `Near' if $\alpha < 8$~kpc  and defined star particles as belonging to the `Far' group if $\alpha > 8$~ kpc.  A similar process of splitting the data into a `Near' group and  a `Far' group was also carried out for orbits in Model A. We note that this is  a very simplistic way to attempt to reproduce the BRC and FRC which does not account for magnitude limits of the observations or extinction effects, nor does it realistically account for a possible dependence on stellar populations, and is therefore only meant to be a crude proxy for the observed velocity distributions. Note that similar,  cuts have been used by others in the comparison of simulations to data \citep{li_shen_12,vasquez_etal_13, debattista_etal_16_fractionation}.

Figure~\ref{fig:Kin_A} shows Model A (top row) and Model C (bottom row) in the HCR frame. The left-most column shows a surface density map of all stars in the two bars from our helio-centric point of view. The asymmetry in the shape of the bulge due to the near-side of the bar being much closer to us is clearly seen and is similar to that observed in the MW \citep{wegg_gerhard_13}.  The second column shows  histograms of \vlos\ for `Near' stars (red) and  `Far' stars  (blue) at $l=0^\circ, b=-6^\circ$ with solid red and blue lines showing the mean velocities of the two distributions obtained by Gaussian fitting. It is clear that the `Near' (red) histogram peaks at  negative \vlos\ values as in the case of the BRC while the  peak of the `Far' (blue) histogram  is at slightly more positive \vlos. The difference in the mean velocities of `Near'  and `Far' side stars is defined as $\Delta V_{\rm los} =  {\langle{V_{\rm los}}\rangle}_{\rm Near} -  {\langle{V_{\rm los}}\rangle}_{\rm Far}$. At $l=0^\circ, b=-6^\circ$, $\Delta V_{\rm los} = 24.3$ \kms (Model A) and $\Delta V_{\rm los} = 28.4$\kms (Model C). The 3rd column in this figure shows the CDFs of these two \vlos\  histograms. The overall behavior of the CDF of the `Near' and `Far' side stars in the third column are qualitatively similar to the BRC and FRC stars observed by \citet{vasquez_etal_13}, (although the separations in our simulations of 24-28\kms are significantly smaller than the observed separation of $\sim$50\kms), perhaps because the strength of the X-shape in our bar is weaker than in the Milky Way).

G14 showed that 2D maps of  $\Delta V_{\rm los}$  reveal kinematic differences between simulations with and without X-shapes. In particular they found that a model with a strong X-shape (e.g. their model B3) showed an asymmetric ring like structure populated with negative $\Delta V_{\rm los}$ similar to that seen in the 4th column of Figure~\ref{fig:Kin_A} surrounding an inner circular region centered on  $l=0^\circ, b=0^\circ$ where $\Delta V_{\rm los} \sim 0$ or slightly positive (note that G14 used a color scheme that is the opposite of that used in this figure). G14 also found that other types of kinematic maps (e.g. showing  \siglos, or galactocentric azimuthal velocities and velocity dispersions) were not able to distinguish between models with and without X-shapes. We also made  maps of \siglos\ and azimuthal velocities for both the full simulations and individual orbits and found them to be uninformative, hence do not include them here.  

%%%%%%% Figure 6 -- Sample Orbits -- %%%%%%%
\begin{figure*}
	\includegraphics[width=.7\textwidth]{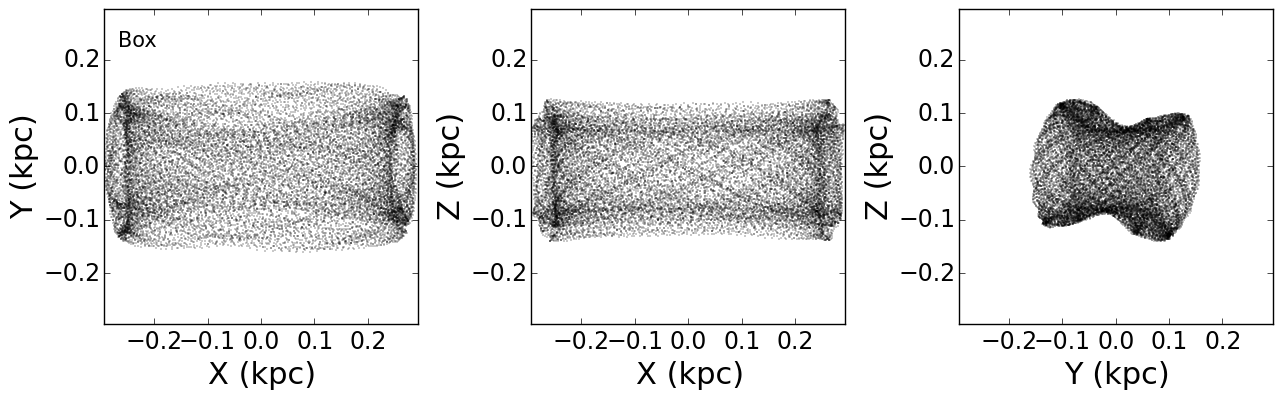}
	\includegraphics[width=.7\textwidth]{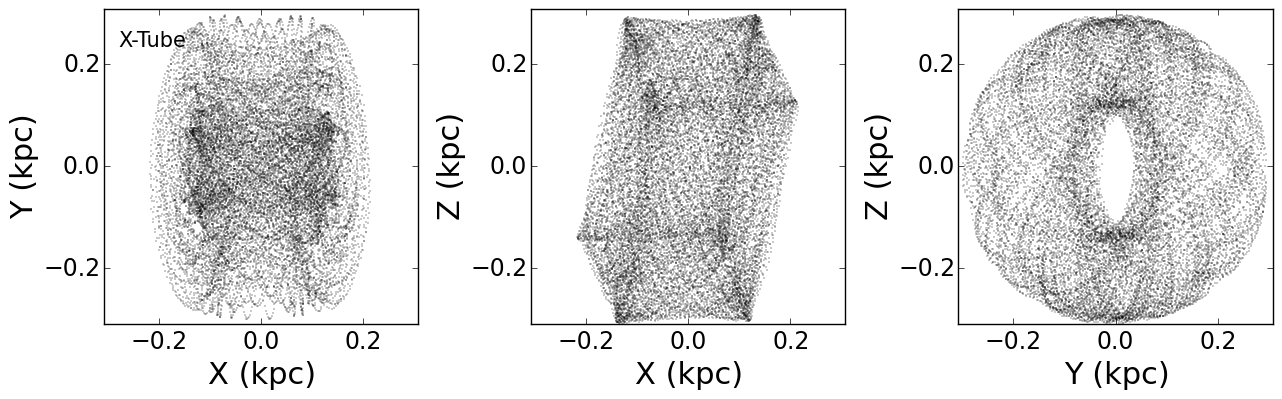}
	\includegraphics[width=.7\textwidth]{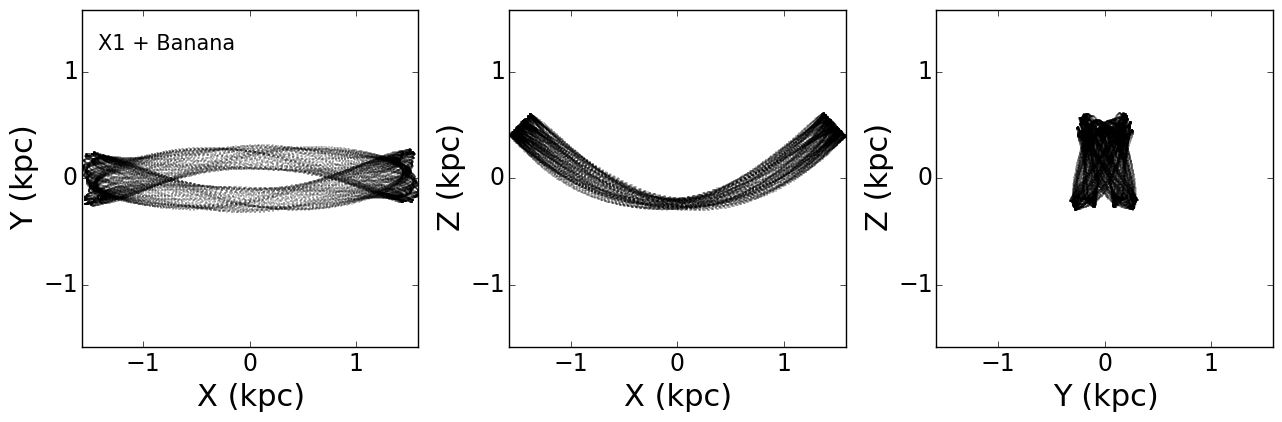}
	\includegraphics[width=.7\textwidth]{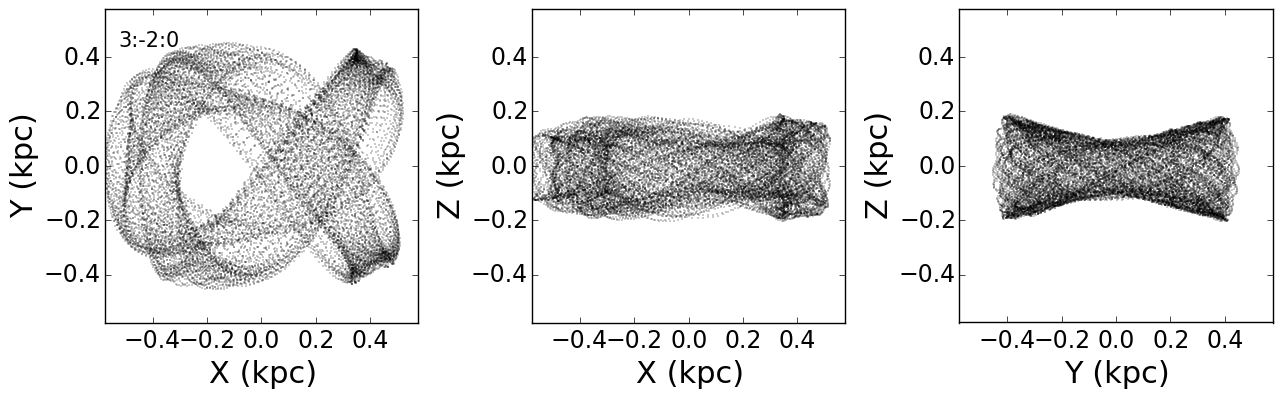}
	\includegraphics[width=.7\textwidth]{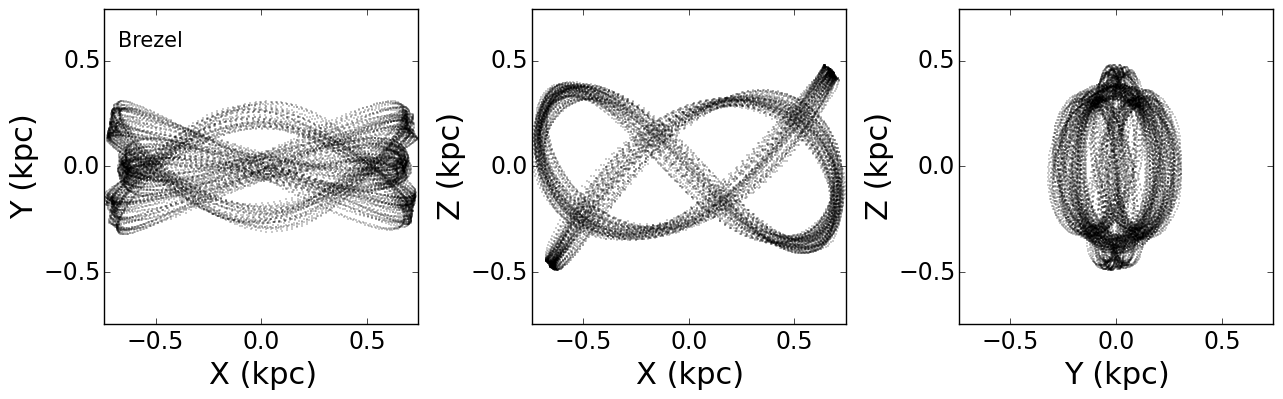}
    \caption{Three different projections of 5 different orbits in Model A. From the top to bottom: a box orbit, an $x$-tube orbit, a banana ( {x1v1}) orbit (2:-2:1), a fish/pretzel orbit (3:-2:0), and a brezel  orbit (3:0:-5).}
     \label{fig:samp_orb}
\end{figure*}
%%%%%%% Figure 6 -- Sample Orbits -- %%%%%%%

Figure~\ref{fig:Kin_A} also shows that there is very little difference between Model A and Model C in either spatial or kinematic distributions. Since the SMBH in Model C (which is about 1.6 times more massive than the SMBH located at the Galactic center) produces no perceptible difference from Model A (without a SMBH) we conclude that a BH of this low a mass does not significantly alter the dynamical structure of the bulge and hence in the sections that follow we confine the discussion to orbits in Model A. We note that  V16 and \citet{brown_etal_13}  analyzed a model with a 10 times more massive SMBH (referred to as Model B) and found that the more massive SMBH does in fact alter the orbital structure in the nuclear region and produces slightly different kinematical signatures. Since this larger black hole mass is not relevant to the MW we do not show results for Model B here.

%%%% Figure 6 -- Orbits that show peanut or X Shape in unsharp masks -- %%%%
\begin{figure*}
% original figures in /Users/mvalluri/Desktop/Current/Bar_simulations/X-shape/unsharp_masking/orbit_edgeon/
% made with program ../Unsharp_orbs.f; k =2 (kernel width), filename has num1_num2 - which are the lower and
% upper limits (lo3, hi3) of the grey scale used to highlight the features in the unsharp mask.
	\includegraphics[trim=0.pt 0.pt 0.pt 0.pt ,clip,width=.48\textwidth]{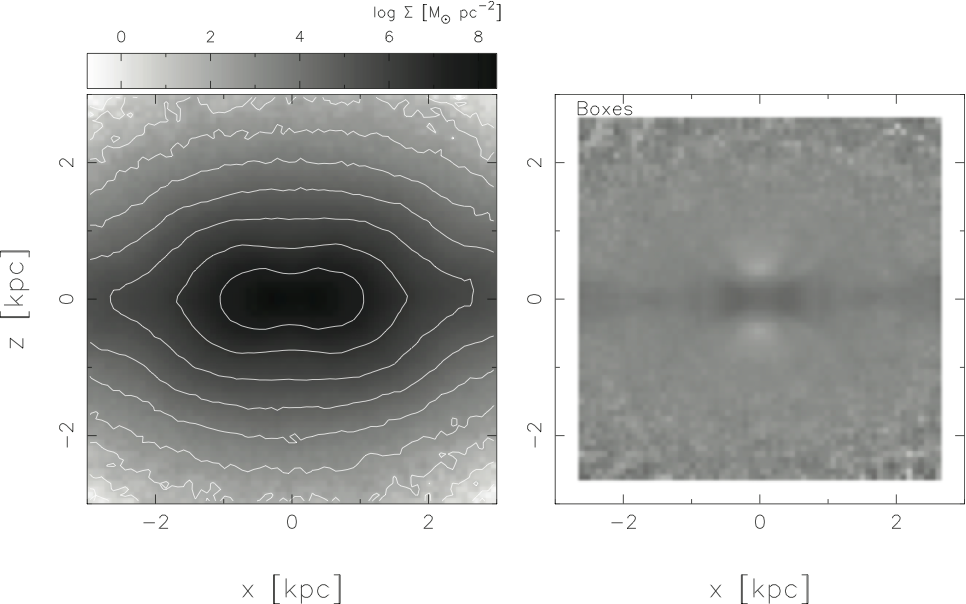}
	%box with k=4, lo3=0.692, hi3=0.694
	\includegraphics[trim=0.pt 0.pt 0.pt 0.pt ,clip,width=.48\textwidth]{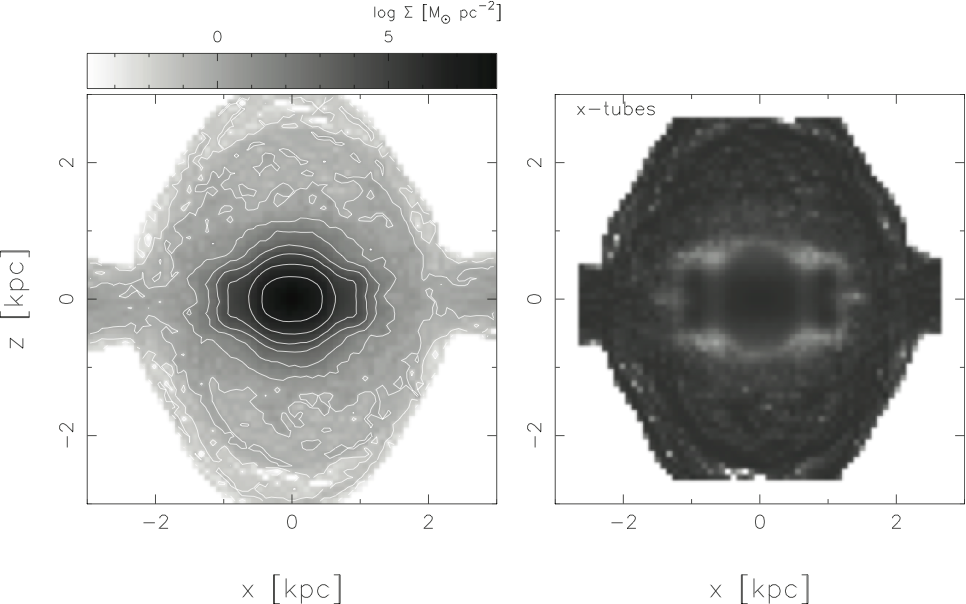}
	%xtub with k=4, lo3=0.684, hi3=0.695
	\includegraphics[trim=0.pt 0.pt 0.pt 0.pt ,clip,width=.48\textwidth]{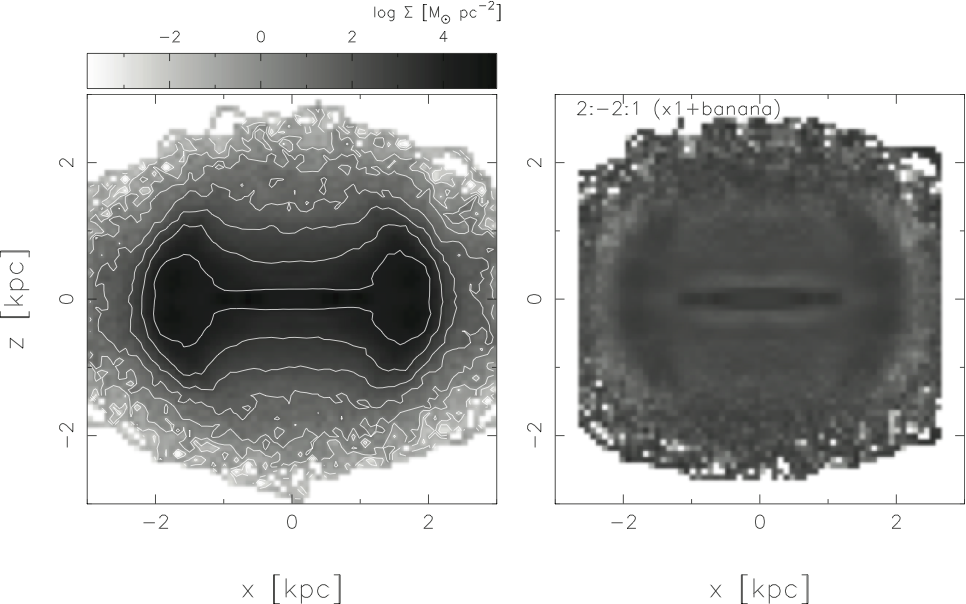}
	%x1ban: k=4, lo3=0.689, hi3=0.694
	\includegraphics[trim=0.pt 0.pt 0.pt 0.pt,clip,width=.48\textwidth]{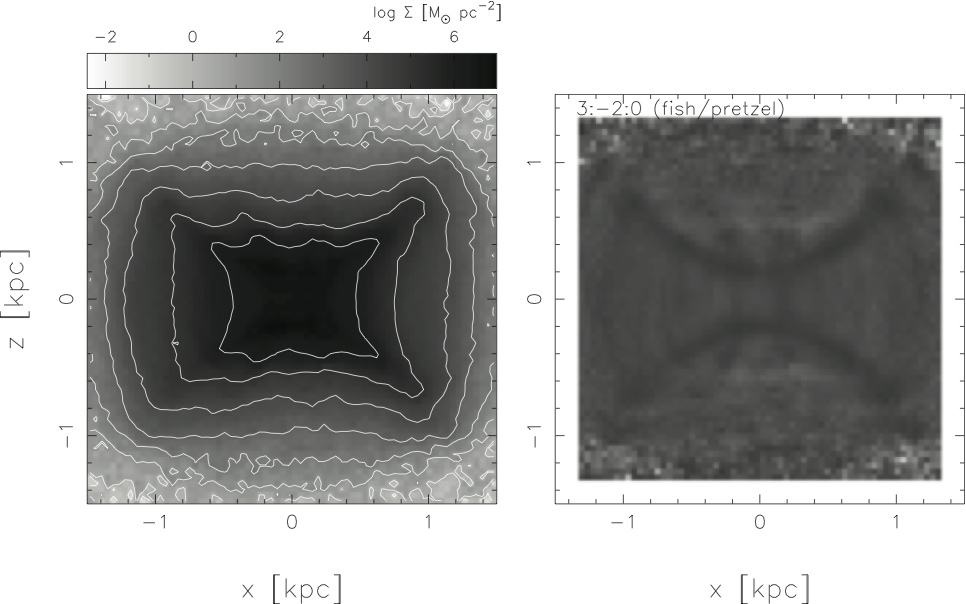}
	% fpz; k=4,lo3=0.69, hi3=0.694
	\includegraphics[trim=0.pt 0.pt 0.pt 0.pt,clip ,width=.48\textwidth]{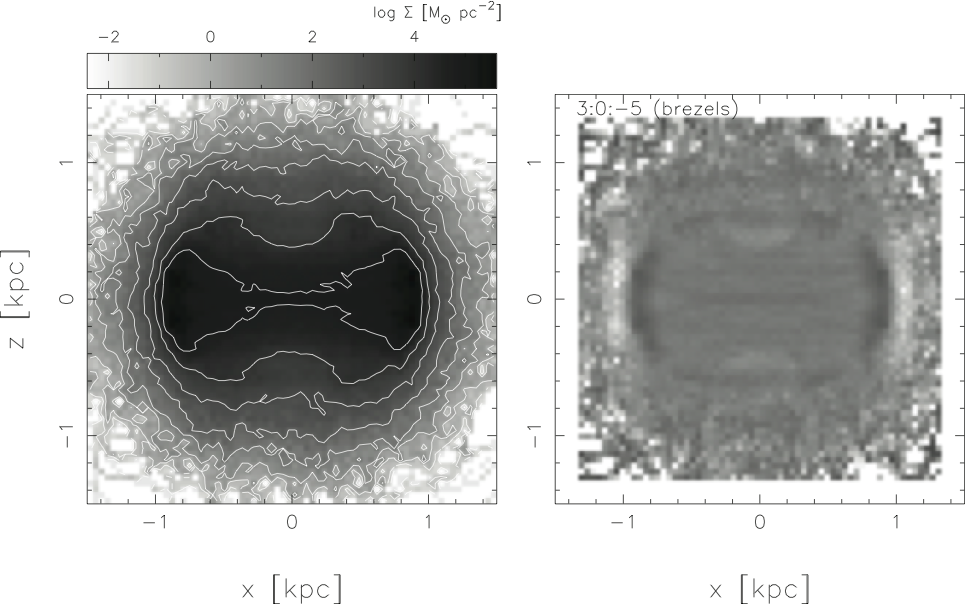}
	% brez: k=4, lo3=0.691, hi3 = 0.694
%	\includegraphics[trim=0.pt 0.pt 0.pt 75.pt ,clip,width=.48\textwidth]{ztubk4_685_695.png}
	\caption{Each pair of panels shows projected surface mass density (left) and unsharp masked image (right) for five orbit families from Model A as labeled. The box orbits (top row left) and $x$-tube orbits  (top row right)  show a distinct peanut shape in projected surface density maps but no  X-shape in unsharp masked images. Resonant 2:-2:1 x1+banana orbits  (2nd row left) show a BP/X-shape in projected surface density map (left) but not in unsharp masked image (right) where it shows two shell-like structures at $\pm 2$~kpc. The resonant boxlet 3:-2:0 (fish/pretzel) orbits show a distinct X-shape both in projected surface density and unsharp masked images (2nd row right). The resonant 3:0:-5 (brezel) orbits show a peanut shape (left) and faint X-shape with two shell-like structures at $\pm 1$~kpc (right) (3rd row).}
\label{fig:x_orb}
\end{figure*}
%%%% Figure 6 -- Orbits that show peanut or X Shape in unsharp masks -- %%%%

\subsection{Orbits}
\label{sec:orbits} 

Orbits of 10,000 randomly selected particles from the self-consistent particle distribution for  Model A were integrated starting from initial conditions corresponding to the positions and velocities of particles in the snapshot at $t=700$ (after the bar growth largely saturates). The orbits were integrated in Cartesian coordinates in the frozen potential of the full simulation, after taking into consideration the appropriate Coriolis and centrifugal pseudo-forces determined by the pattern frequency $\Omega_p$ of the bar.   The pattern frequency of the bar in Model A in program units is 0.117, which for our choice of physical units gives a pattern frequency $\Omega_{p} = 47.98$\kmskpc. The most recent estimate \citep{portail_etal_16} of the pattern frequency of the Milky Way bar is $\Omega_{p}= 39.0\pm 3.5$\kmskpc \, (for a bar length  { $5.3\pm0.36$kpc). These differences in pattern frequency (as well as differences in the strength of the bar) could account for some of the differences between the observations and the model seen in Figure~\ref{fig:Bra_fitAB}.}

All orbits were integrated for 1000 time units (equal to  3.69~Gyr) in the rotating frame and saved at 20,000 equally spaced time intervals. Each orbit was then analyzed using a spectral analysis code \citep{valluri_merritt_98,valluri_etal_10} and the fundamental orbital frequencies were used to classify orbits using the scheme described in V16. 

As discussed by  V16, orbits in N-body bars are essentially identical to orbits in triaxial ellipsoids, with the main difference being that they are modified by the pseudo-forces arising from figure rotation. Bar orbits belong to the same five main families found in triaxial ellipsoids: boxes, short-axis ($z$) tubes, inner and outer long-axis ($x$) tubes and chaotic orbits. V16 showed that $\sim$ 60\% of bar orbits in two bar simulations were box orbits that are parented by the linear orbit that oscillates along the $x$-axis (1st row of Fig.~\ref{fig:samp_orb}). They also found a small fraction (8.5\%) of long-axis ($x$)-tube orbits (2nd row of Fig.~\ref{fig:samp_orb}). The orbit traditionally referred to as the ``prograde x1 orbit", and its vertical bifurcation the  2:-2:1 banana orbit  is shown in 3rd row of Fig.~\ref{fig:samp_orb} (this family comprises 3\% of all bar orbits). V16 also identified higher order resonant ``boxlets'', members of the box orbit family associated with the 3:-2:0 resonance and the 3:0:-5 resonance. The former, which we refer to as the ``fish/pretzel'' resonance (although it is different from the 3:0:-2 fish and  4:3:0 pretzel families found in stationary triaxial potentials) is shown in the 4th row of Fig.~\ref{fig:samp_orb} and is the largest resonant family, comprising $\sim$6\% of bar orbits. Less than 2\% of orbits  were associated with the 3:0:-5 ``brezel'' resonance (5th row of Fig.~\ref{fig:samp_orb}) which was proposed by \citet{portail_etal_15b} as the back-bone of the X-shape. 

A small number of orbits in Model A are retrograde short-axis tube orbits (which are elongated along the $y$-axis of the model and resemble x4 orbits) and none of the orbits in this model were associated with prograde x2 orbits.  Our analysis of short axis tubes reveals that they show neither a BP shape nor an X-shape in projection and do not have line-of-sight density or kinematical distribution consistent with MW observations, hence for the rest of this paper this family is ignored. As mentioned previously we also  ignore chaotic orbits, since their shapes evolve with time, although at 18-20\% of bar orbits they are not an insignificant population.

\section{Results}
\label{sec:results}

\subsection{Projected distributions and unsharp masking for orbits}
\label{sec:proj_orbs}

With the orbit classifications from V16 in hand, we set about analyzing each orbital family to determine which family (or families) is (are) primarily responsible for producing the BP/X-shape  {in this model}. This was done by co-adding all orbits belonging to a particular family and constructing the projected density distribution for that family. We are justified in superposing orbits in this manner because all orbits were integrated for the same amount of time and saved at the same time-steps and hence orbit-superposition is akin to considering a large number of stars (on a given orbit) whose positions reflect the density distribution along the orbit. We present the projected X-shape only in the edge-on bar since it is  most clearly observed in this orientation (but similar results were obtained in the HCR frame). The HCR projection is used for examining line-of-sight number counts and line-of-sight kinematics since we compare these metrics with observations for the MW.

As we saw in Figure~\ref{fig:unsharpfullsims}, the projected density distribution of the full simulation shows the box/peanut (BP) shape while the process of unsharp masking best reveals the presence of the X-shape.  We now make similar plots for each of the 5 different families, by co-adding all members of a given family (Figure~\ref{fig:x_orb}). Each row shows co-added plots for two orbit families and for each family we show the projected surface density in the left panel and the unsharp masked image on the right.   

 The box orbits (top row, 2 left panels) show a distinct peanut shape and even a hint of an X-shape in the projected surface density plot but does not show an X-shape in the unsharp masked image. 
 
 The $x$-tubes (top row, 2 right panels) shows a small scale peanut shaped structure (within $\pm 1$~kpc) in both the full projection and in the unsharp masked image. This smaller scale peanut (along with the larger peanut produced by the boxes) might contribute to the nested peanuts recently observed by \citet{ciambur_graham_16}. 
 
The x1+banana orbits (2:-2:1 resonance) clearly shows a peanut shape, with a hint of a broad X-shape, in the projected surface density plot but the unsharp masked image shows no evidence of an X-shape. Rather, in the unsharp masked image we see two shell like structures perpendicular to the disk at $\pm 2$~kpc. 

The two resonant boxlet families, the 3:-2:0 fish/pretzel (2nd row right) and the  3:0:-5 brezel (3rd row) show distinct X-shapes both in the projected surface density plots and in the unsharp masked images. In the case of the brezels the X-shape is much broader and more diffuse and it also shows the double shell-like structure at $\sim \pm1$~kpc similar to that seen for the x1+bananas.  
 
It appears from the projected density plots that several families contribute to the box/peanut shape (boxes, $x$-tubes, x1+bananas) and the X-shape (fish/pretzel and brezels).  Recent analysis of x1+banana orbits by \citet{portail_etal_15b} and \citet{Qin_Yujing_MS} shows that although this family is frequently invoked as causing the X-shape these orbits do not appear except towards the ends of the bar and hence are probably not a significant contributor to the X-shape in the inner half of the bar. Furthermore,  \citet{Qin_Yujing_MS}, who analyzed the MW bar model presented by \citet{shen_etal_10}, and classified orbits using a method based on orbital angular-momentum finds results completely consistent with those presented here. He compared the density images of the stacked orbital families to argue that ``x1-like" orbits (non-resonant but closer to the the x1 resonant family) and ``boxy orbits" (box orbits showing larger deviations from x1 resonant family) contribute to different parts of peanut/X-shape. 

In unsharp masked images however, only fish/pretzels  show a distinct, thin X-shape, while brezels show a more fuzzy X-shape. Although the unsharp masked images suggests that fish/pretzel orbits and brezel orbits contribute significantly, both families also contribute a significant fraction of their total mass outside the X-shape.

%From unsharp masking, \citet{li_shen_12} conclude that about 7.\% of the mass of the bar is associated with the X-shape.  In contrast however, \citet{portail_etal_15a} used a different method to determine the fraction of stellar mass associated with the X-shape: they constructed a family of ellipsoidal isodensity surfaces with each ellipsoid required to fit inside the true 3D isodensity surfaces and estimated to total mass associated with the non-ellipsoidal (BP/X-shaped) component to be $24^{+5}_{-4}$\% of the total bulge mass.  {Furthermore \citet{portail_etal_15b} from an analysis of orbits, that 40 - 45\% of orbits in their MW bulge model are associated with the BP/X shape.}

The non-resonant box orbits (which constitute 63\% of the mass of the bar), x1+banana orbits (3\%), fish-pretzels (6\%)  and brezels (1.5\%) all show a distinct BP/X shape in projected surface density images, while only fish-pretzels and brezels (7.5\% in total) show the X-shape in unsharp masked images. This suggests that unsharp masking may underestimate the mass associated with the X-shape and that in fact the majority of bar orbits contribute to the BP/X shape.

In the next section we see further evidence that the majority of orbits (especially the non-resonant box family) contribute to stellar number counts that provide evidence for a 3 dimensional BP/X-shaped bulge in the MW. 

\subsection{3D ``excess-mass'' distribution}
\label{sec:mod_xmass}
\begin{table}
	\centering
	\caption{Fraction of mass outside ellipsoids/orbits contributing to X-shape.}
	\label{tab:orbfracs}
	\begin{tabular}{lc} % 
		\hline
%Orbit               &          Fraction of mass  &     Fraction of orbits\\
%                    &       outside ellipsoids    &      contributing to X\\
Orbit               &          Fraction of mass  \\
                    &       outside ellipsoids    \\
\hline
Model A            &      0.23       \\% &  --- \\
Model C            &      0.19      \\%  &   --- \\                          
Boxes              &      0.25        \\%&     0.82  \\
Brezels            &      0.29       \\% &     0.69  \\
Fish/Pretzels      &      0.09     \\%   &     0.44  \\
x1+banana          &      0.33     \\%   &     0.90  \\
X-tubes            &      0.10        \\%&     0.69   \\
%Z-tubes           &      0.05      \\%  &     0.45\\
%Chaotic           &      0.22      \%  &     0.95\\
\hline
	\end{tabular}
\end{table}
%%%% Figure xx -- Orbits - excess mass -- %%%%
\begin{figure*}
	\includegraphics[trim=0.pt 0.pt 0.pt 0.pt ,clip,width=.3\textwidth]{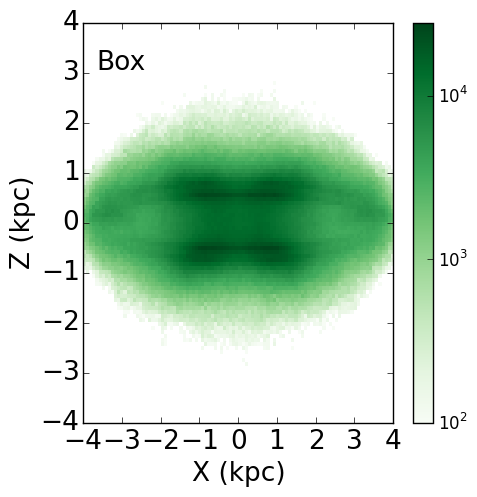}
	%box 
	\includegraphics[trim=0.pt 0.pt 0.pt 0.pt ,clip,width=.3\textwidth]{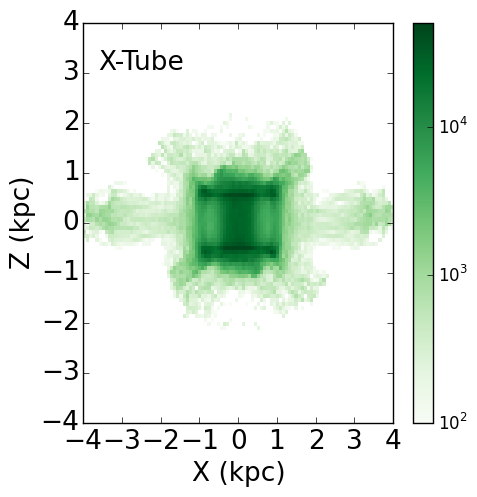}
	%xtub 
	\includegraphics[trim=0.pt 0.pt 0.pt 0.pt ,clip,width=.3\textwidth]{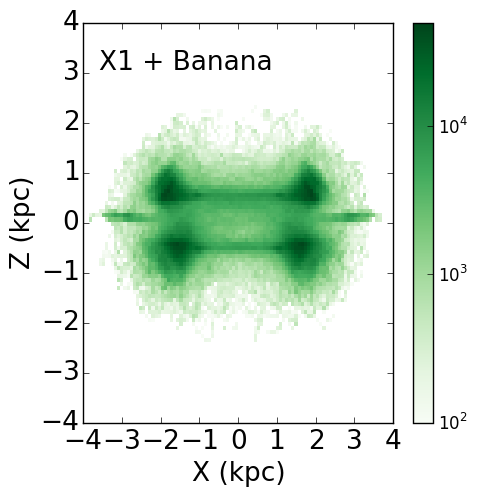}
	%x1ban: 
	\includegraphics[trim=0.pt 0.pt 0.pt 0.pt,clip,width=.3\textwidth]{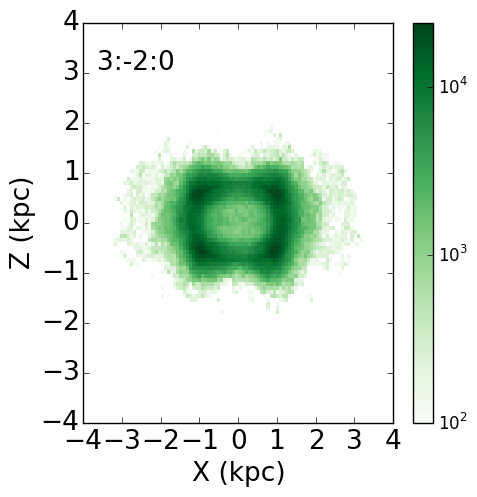}
	% fpz; 
	\includegraphics[trim=0.pt 0.pt 0.pt 0.pt,clip ,width=.3\textwidth]{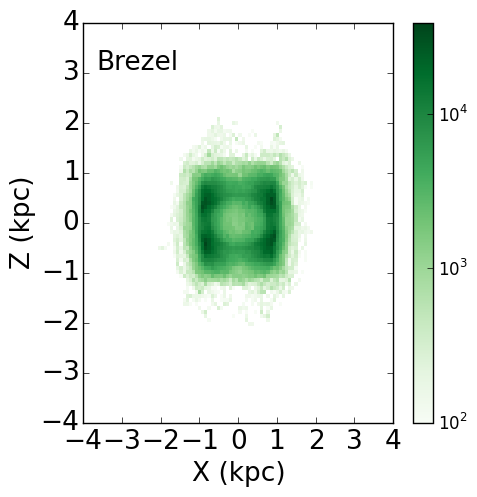}
	% brez: 
	\caption{Projected excess mass density on the $x-z$ plane for each of the five orbit families from Model A as labeled. The color gradient as shown in the bar on the right of each plot is in units of $\msun\ / kpc^{2}$. The box orbits (top row left) and resonant 2:-2:1 x1+banana orbits  (top row right) show a clear BP/X-shape at large radii. The $x$-tube orbits (top row middle) show a hint of an X-shape but at very low density levels. Resonant boxlet 3:-2:0 fish/pretzel orbits (bottom row left) and resonant 3:0:-5 brezel orbits (bottom row right) show a  BP/X-shape at small radii ($\leq 2$~kpc).}
\label{fig:orbits_Xmass}
\end{figure*}
%%%% Figure xx -- Orbits excess mass -- %%%%

 Using the method described in Section~\ref{sec:methods}  we also computed the excess mass outside ellipsoids for each co-added orbit family in a manner similar to that carried out for the full model. Figure~\ref{fig:orbits_Xmass} shows the excess mass outside ellipsoids for each co-added orbit family. %The ellipsoids for each orbit are determined for each co-added orbit family at the same density levels as from Model A. 
When these maps are compared with Figure~\ref{fig:model_Xmass} for the full simulations it is clear that except for the $x$-tubes (top row middle panel) which shows only a hint of a BP/X-shape, all other orbit families   show a clear BP/X shape suggesting that they all contribute to the 3D excess mass outside ellipsoids. The excess mass  for each co-added orbit family is given in the 2nd column of Table~\ref{tab:orbfracs} and shows that all  families in bar Model A  contribute some fraction of  their total  mass to the BP/X-shaped region. Furthermore Figure~\ref{fig:orbits_Xmass} shows that each family contributes to the BP/X shape over a specific radial range: the resonant brezel and fish/pretzel orbits contribute closest to the center of the bar within $|x| < 1.5$~kpc, the boxes contribute most significantly at intermediate radii, while the x1+banana orbits contribute primarily at the ends of the bar. In other words the BP/X shape {\it is the bar} and the vast majority of orbits contribute to it.

\subsection{Bimodal distribution in heliocentric distances}
\label{sec:bifurcation_orb}
%%%% Table 2: Bifurcation in stellar number counts along  4 lines-of-sight %%%
\begin{table}
	\centering
	\caption{Bifurcation in stellar number counts along four lines-of-sight: distance between peaks}
	\label{tab:rhelio_sep}
	\begin{tabular}{lcccc} % 
		\hline
		   & $b=-4^\circ$ & $b=-5^\circ$ & $b=-6^\circ$ & $b=-7^\circ$  \\
		   & (kpc)             & (kpc)            & (kpc)            & (kpc)            \\
		\hline
		Model A (full) &  0.5 & 1.0 & 1.8 &  2.4 \\
		Boxes (63\%)           &  0.8 & 1.6 & 1.9 &  2.1  \\
		X-tube (8.5\%)          &  0.1 & 0.1 &  0.1 &	0 \\
		x1+banana (3\%)   & 3.    & 3.2  & 3.3  & 3.7 \\
		3:-2:0  (5.9\%)          & 1.5  & 1.6 & 1.7  & 1.8  \\
		Brezel  (1.4\%)         & 1.2  & 1.2  & 1.3  & 1.4 \\
		\hline
	\end{tabular}
\end{table}
%%%% Table 2: Bifurcation in stellar number counts along  4 lines-of-sight %%%

As discussed in the introduction, one of the first signatures of the three dimensional nature of the BP/X shape in the Milky Way bulge was the detection of  a bifurcation in red clump star number counts as a function of magnitude \citep{mcwilliam_zoccali_10,nataf_10} for $|b| \gtrsim 5^\circ$. This bifurcation has been confirmed observationally by others \citep[e.g.][]{saito_etal_11} and is also found in simulations \citep[e.g.][]{li_shen_12,debattista_etal_16_fractionation}.

We  now examine the distributions of particles along various lines-of-sight for the full simulation and compare them with similar distributions for co-added orbits. To ensure that disk particles are excluded we applied a cut that selects only particles that satisfy the constraint $(x/5)^2 + (y/1.5)^2 <1$  (assuming the semi-major axis length of the bar is 5~kpc and the semi axis length in the disk plane is 1.5~kpc). No cut was applied perpendicular to the disk. All particles in a pencil beam with $\pm0.25$~kpc square cross-section were selected along four different lines-of-sight, at $l=0^\circ$ and $b= -4^\circ, -5^\circ, -6^\circ, -7^\circ$ (similar plots were obtained for $b= +4^\circ, +5^\circ, +6^\circ, +7^\circ$ but are not shown). The heliocentric distance distribution of particles in the HCR frame were fitted using a Gaussian Mixture Modeling (GMM) code (we use the Python code sklearn.mixture.GMM available at \url{http://scikit-learn.org}  which fits at most 2 Gaussian distributions to the distances of star particles along each line-of-sight. The code uses maximum-likelihood to estimate the parameters of the Gaussians.
 
\begin{figure*}
%\begin{minipage}{4.875in}radhist_new/
%%%% Figure 7 -- Bimodal distribution of stars along the line-of-sight -- %%%%
%full simulation
	\includegraphics[width=.21\textwidth]{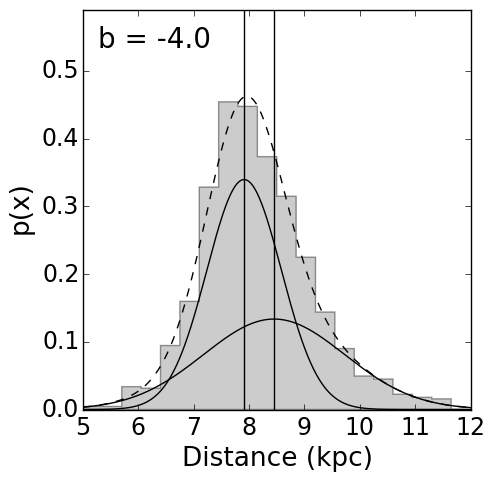}
	\includegraphics[width=.21\textwidth]{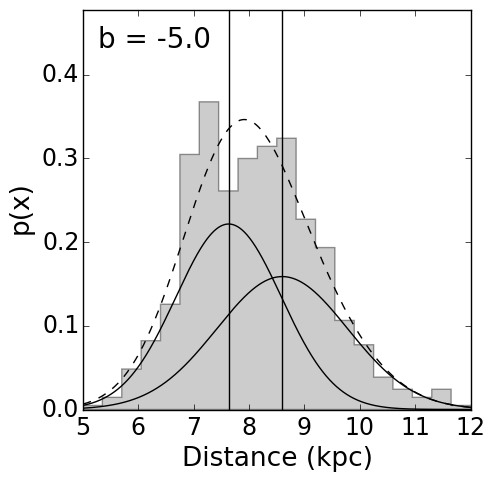} 
	\includegraphics[width=.21\textwidth]{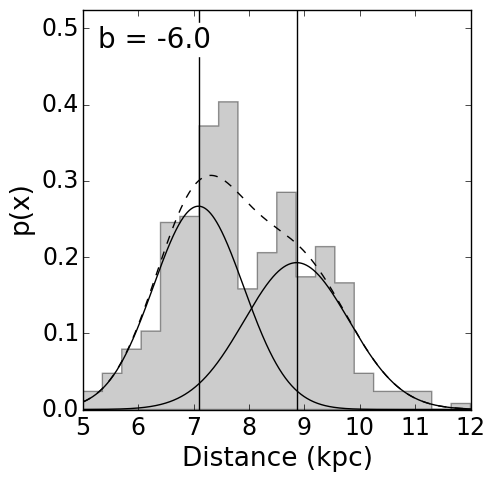}
	\includegraphics[width=.21\textwidth]{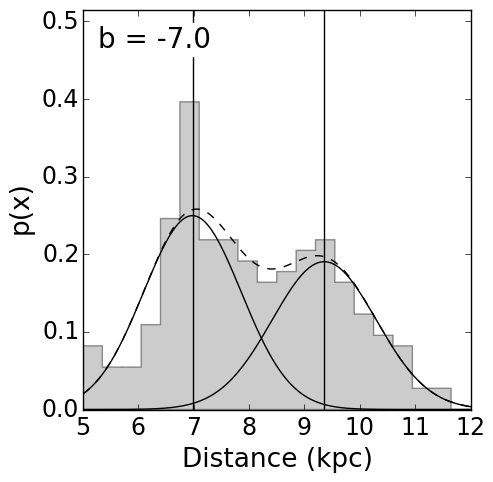}
%boxes
	\includegraphics[width=.21\textwidth]{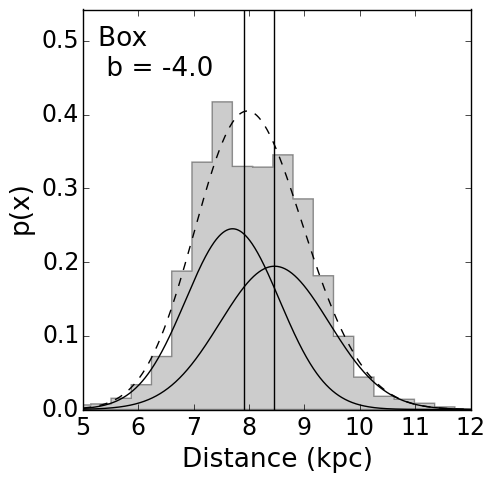}
	\includegraphics[width=.21\textwidth]{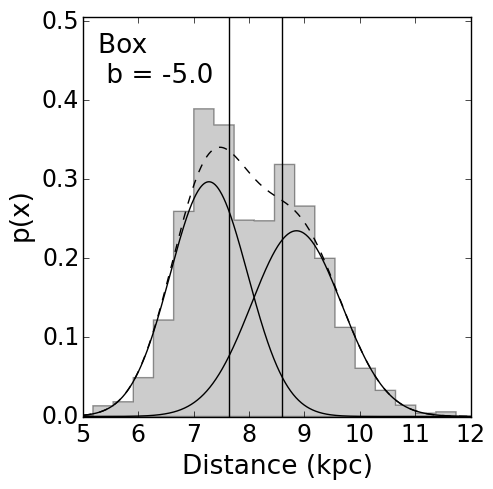}
	\includegraphics[width=.21\textwidth]{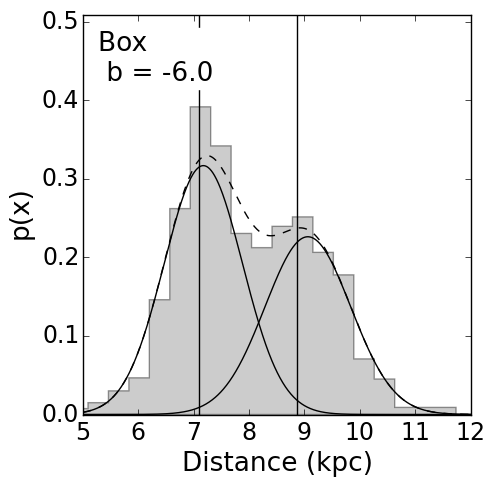}
	\includegraphics[width=.21\textwidth]{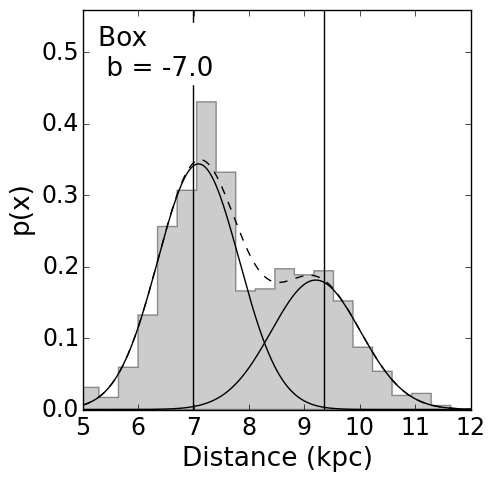}
% x-tube
	\includegraphics[width=.21\textwidth]{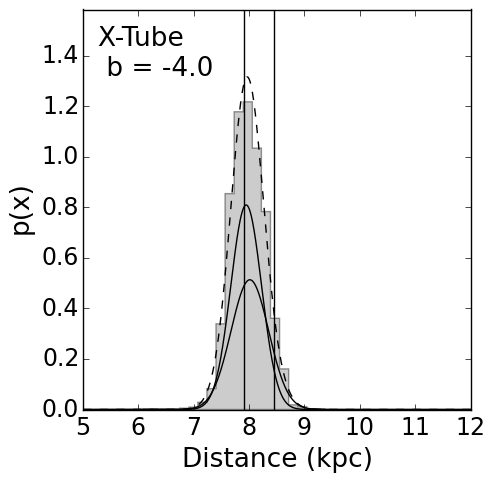}
	\includegraphics[width=.21\textwidth]{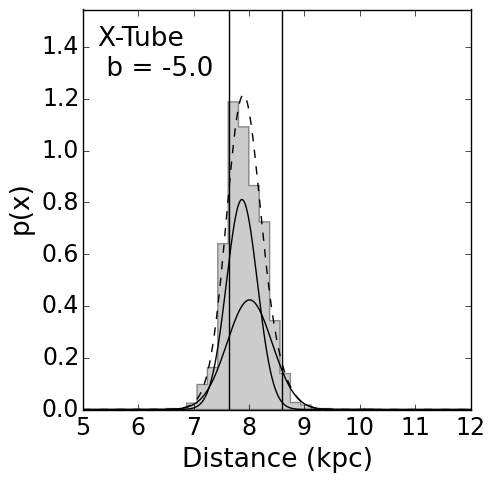}
	\includegraphics[width=.21\textwidth]{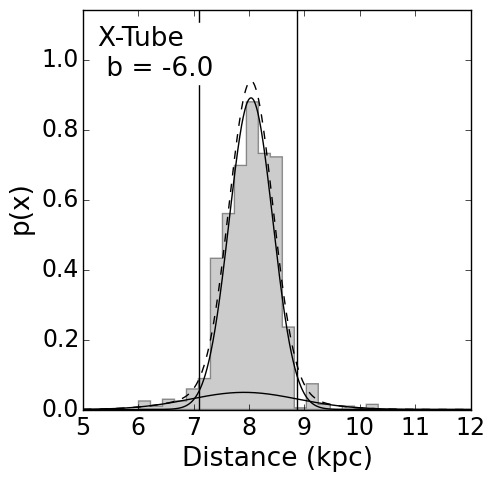}
	\includegraphics[width=.21\textwidth]{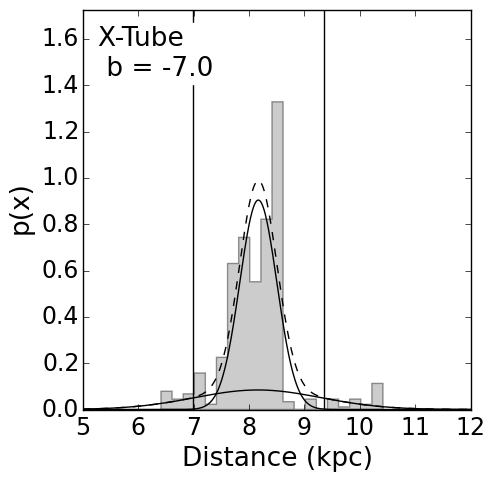}	
% x1ban
	\includegraphics[width=.21\textwidth]{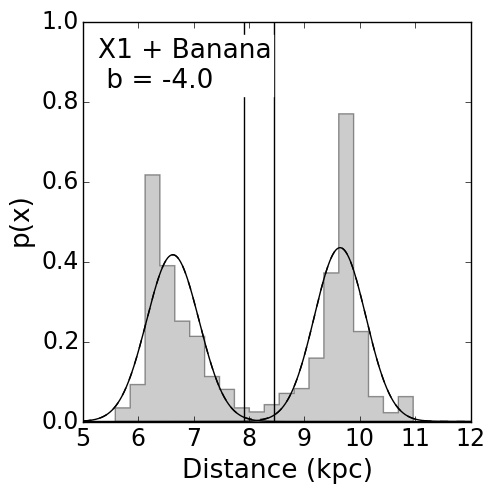}
	\includegraphics[width=.21\textwidth]{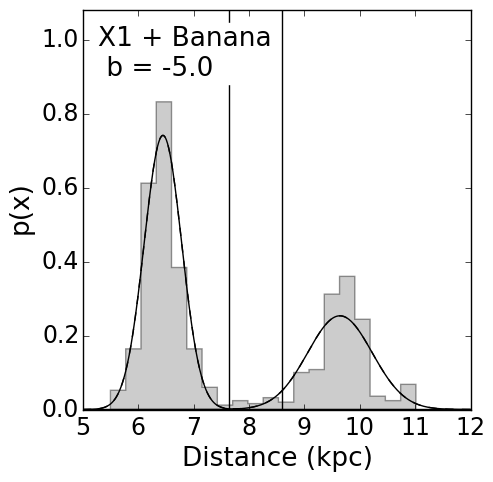}
	\includegraphics[width=.21\textwidth]{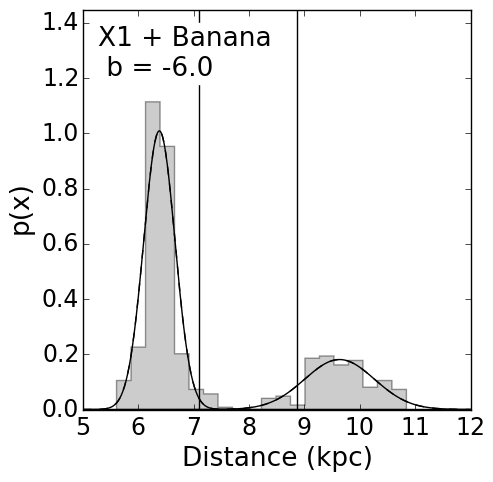}
	\includegraphics[width=.21\textwidth]{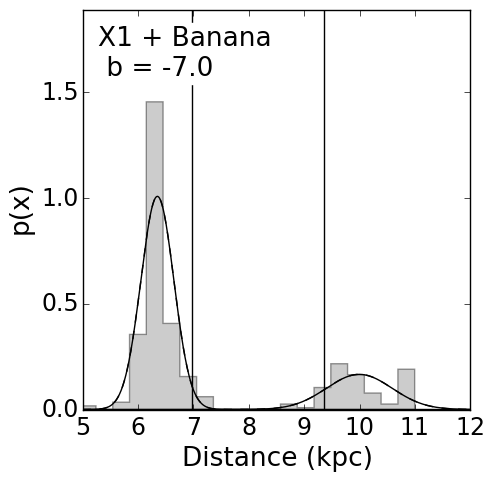}
% fish
	\includegraphics[width=.21\textwidth]{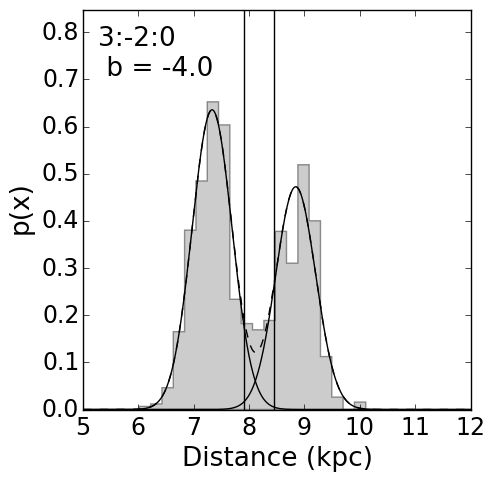}
	\includegraphics[width=.21\textwidth]{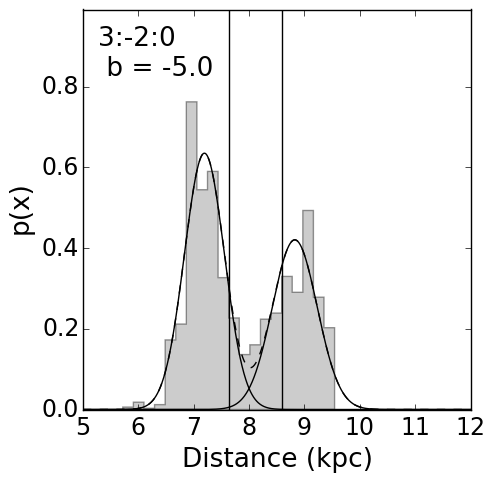}
	\includegraphics[width=.21\textwidth]{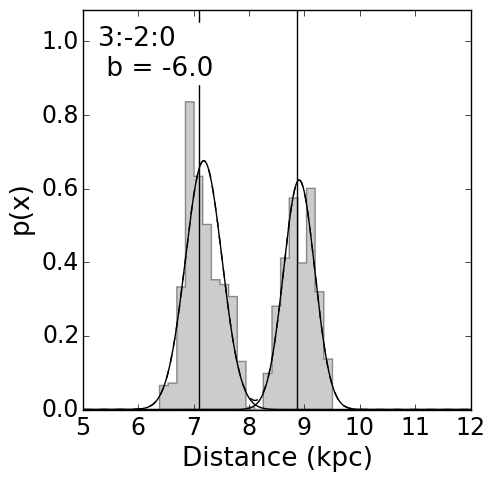}
	\includegraphics[width=.21\textwidth]{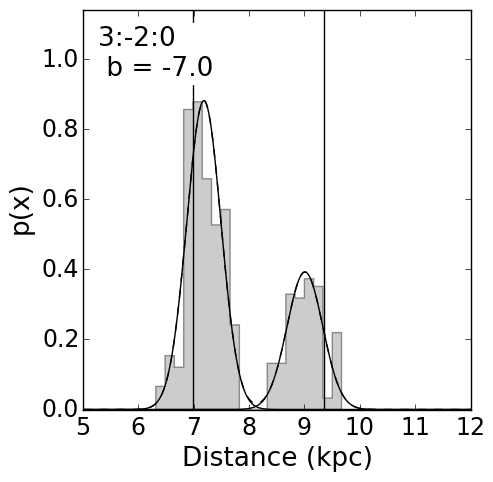}
% brez
	\includegraphics[width=.21\textwidth]{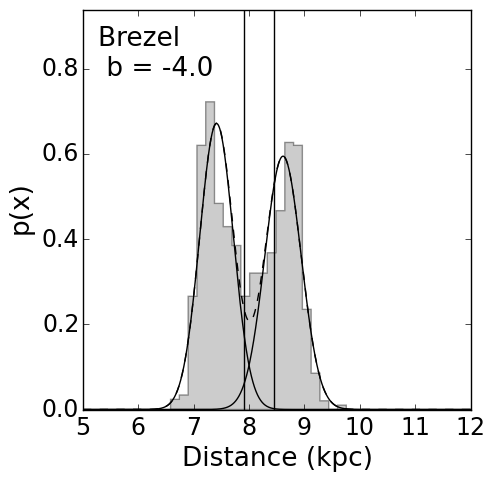}
	\includegraphics[width=.21\textwidth]{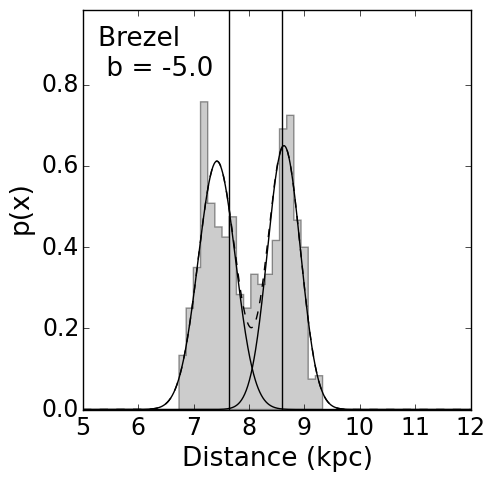}
	\includegraphics[width=.21\textwidth]{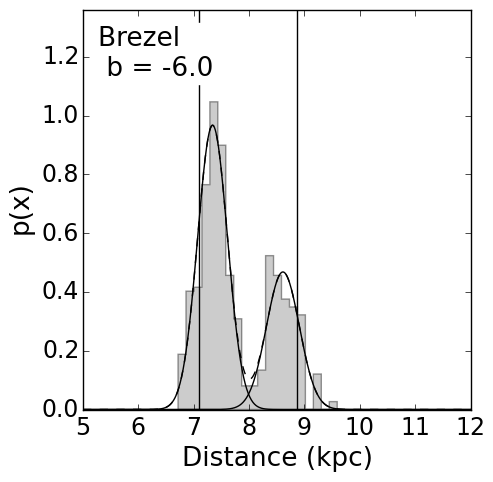}
	\includegraphics[width=.21\textwidth]{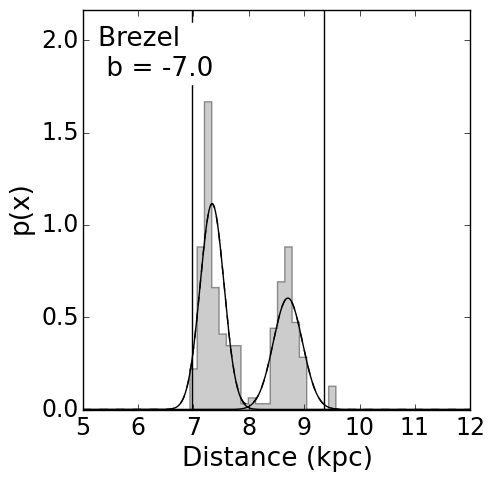}	
	
%\end{minipage}	
%\begin{minipage}
%\begin{center}
%\end{center}
%\end{minipage}	
 \caption{Normalized histograms showing number of stars along the line-of-sight as a function of heliocentric distance at $l=0^{\circ}$ and 4 different $b$ values as indicated in the legends for the full simulation of Model A (top row), and five different orbit families also from Model A co-added (subsequent rows). Solid curves show the two Gaussians (see text for details) that best fit the distribution, dashed curve shows the sum of the Gaussians. Vertical lines in the top row mark the locations of the means of the two Gaussians and these lines are reproduced in panels below. Only the peaks of the Gaussians for box orbits (2nd row) come close to matching the locations of the vertical lines.
    \label{fig:radhist}}
\end{figure*}

In Figure~\ref{fig:radhist} the grey histograms show the number of star particles as a function of  heliocentric distance along various lines-of-sight. The two solid curves show the Gaussians obtained from fitting the distribution of particles with the GMM code. The dashed curve shows the sum of the two Gaussians.  The top row shows the distributions along four lines-of-sight for the full simulation: the locations of the means of the GMM Gaussians are marked with vertical lines, which are reproduced for reference at  the same locations in the  plots for co-added orbits in the next five rows. For the full simulation the histogram at $l=0^\circ, b=-4^\circ$  is skewed with a single peak, similar to that observed in the Milky Way. The GMM code fits the skewed distribution with two Gaussians with means separated by only 0.5~kpc but with very different standard deviations. For $b \leq -5^\circ$ the bimodality is clearly visible in the histograms. Table~\ref{tab:rhelio_sep} gives the separations between the means of the two Gaussians  (from the GMM code) in kpc for the full simulation (top row) and each of the five co-added orbit families. 

In the full simulation (top row of Figure~\ref{fig:radhist} and Table~\ref{tab:rhelio_sep}) we see that the separation of the peaks increases with increasing $|b|$ in a manner similar to that observed in the MW.  This increase in the separation of the peaks in the number counts is regarded as evidence that our line-of-sight through the bar/bulge is passing through ``two opposing arms'' of the X-shape that get farther apart as  $|b|$ increases.  

%%%% Figure 7 -- Bimodal distribution of stars along the line-of-sight -- %%%%

 {The $x$-tubes in the 3rd row  of Figure~\ref{fig:radhist} show no bimodality at any value of $b$, clearly indicating that they do not contribute to the 3D BP/X-shape (although they are probably contribute to the single peak at $b=-4$).  Recall that $x$-tubes showed a clear peanut shape in projected distribution in Figure~\ref{fig:x_orb}.}

Table~\ref{tab:rhelio_sep}  shows that only the box orbit family shows an increasing separation of peaks with increasing $|b|$ that closely matches the increasing separation of peaks seen in the full simulation.  Furthermore, we see in Figure~\ref{fig:radhist} that only the peaks of the Gaussians for box orbits (2nd row) match the locations of the vertical lines (corresponding to the means of the Gaussians in the full simulation). In contrast, the resonant orbit families: 2:-2:1 x1+banana orbits  { (4th  row of Fig.~\ref{fig:radhist})}, 3:-2:0 fish/pretzel (5th row) and  3:0:-5 brezels (6th row) show nearly constant separation in the peaks at all $b$ values. In the case of the x1+banana and brezels this is probably due to the shell-like structures seen in unsharp masked images for these two families in Figure~\ref{fig:x_orb}. For the 3:-2:0 fish/pretzel orbits (which show a slight increase in separation), it is probably due to the high degree of concavity of the arms of the X-shape. We  also see from Table~\ref{tab:rhelio_sep} that the separation of the peaks in the case of the x1+bananas is much larger at all $b$ values than observed in the full simulation, ruling out the possibility that this family is the dominant contributor to the bifurcation in stellar number counts. Nonetheless, the x1+banana orbits definitely contribute to the tails of the distribution in the full simulation since the distributions in the other orbit families fall off more rapidly at small and large heliocentric distance, than they do in the full simulation.

 The conclusion that can be drawn from the analysis of the bimodal distribution of stars along various lines of sight is that stars on non-resonant box orbits, which are the dominant population in the bar, adequately account for both the bimodal distribution in RC stars, and their increasing separation with increasing $|b|$. Since the resonant boxlet families show a bimodal distribution in heliocentric distance that is nearly independent of $|b|$ none of these families individually can explain the observations, although they contribute to the bimodality.  In fact it appears that the combination of non-resonant and resonant orbit families collectively produce the observed bimodal distribution in RC stars.

\subsection{Line-of-sight kinematics of orbits}
\label{sec:kin_orbs}

%%%% Figure 8 -- Histograms, CDF and 2D velocity distribution maps %%%%%
\begin{figure*}
%\begin{minipage}{4.875in}
\begin{flushleft}
	\includegraphics[width=.21\textwidth]{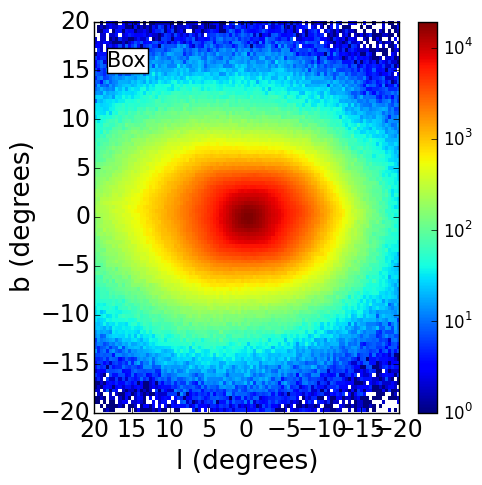}
	\includegraphics[width=.2\textwidth]{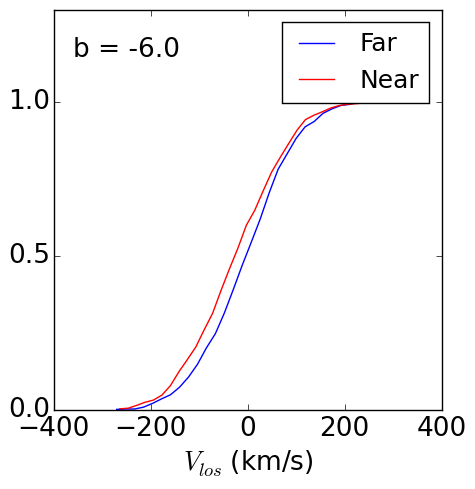}
	\includegraphics[width=.23\textwidth]{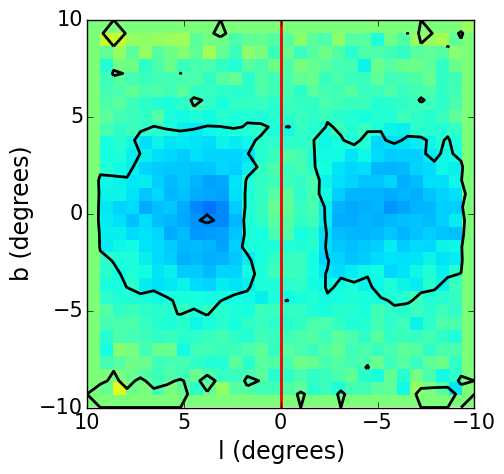}\\
%	\hspace{4cm}
	\includegraphics[width=.21\textwidth]{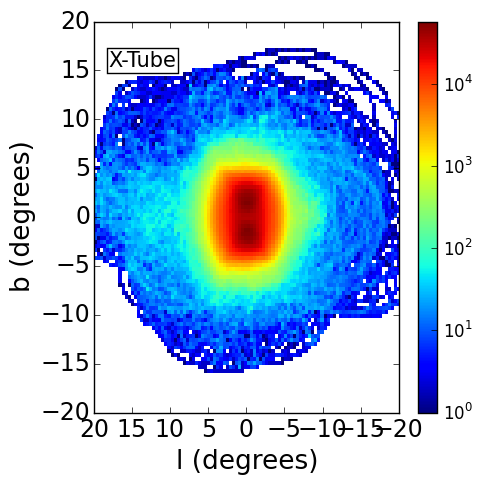}
	\includegraphics[width=.2\textwidth]{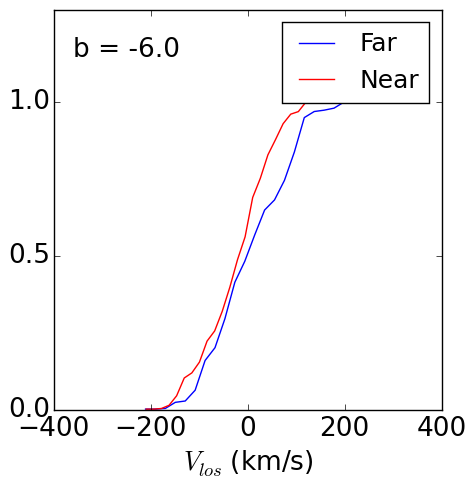}
	\includegraphics[width=.23\textwidth]{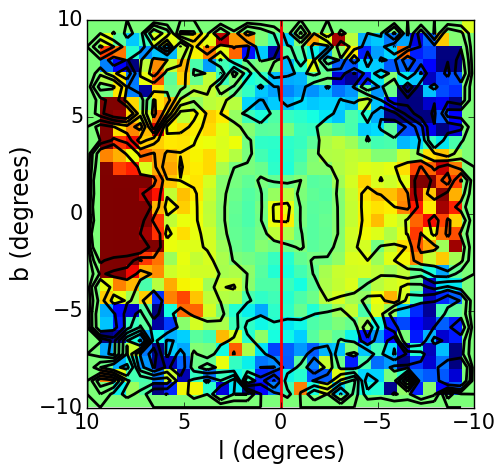}\\
%	\hspace{2cm}
	\includegraphics[width=.21\textwidth]{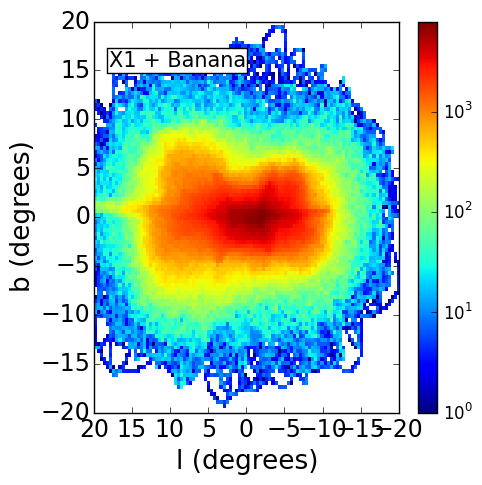}
	\includegraphics[width=.2\textwidth]{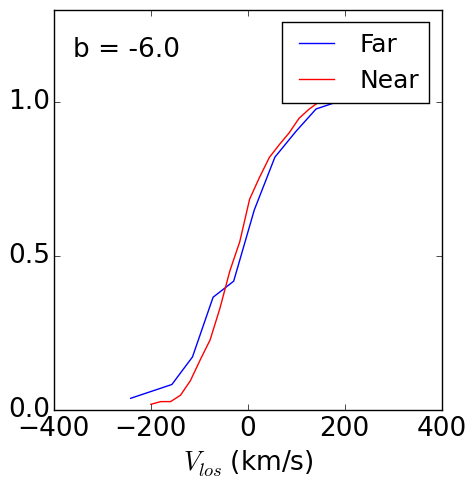}
	\includegraphics[width=.23\textwidth]{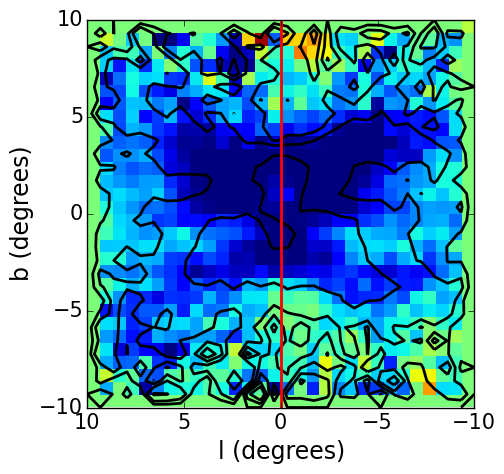}
	\includegraphics[width=.3\textwidth]{gard_7_01_10_max.png}\\
	\includegraphics[width=.21\textwidth]{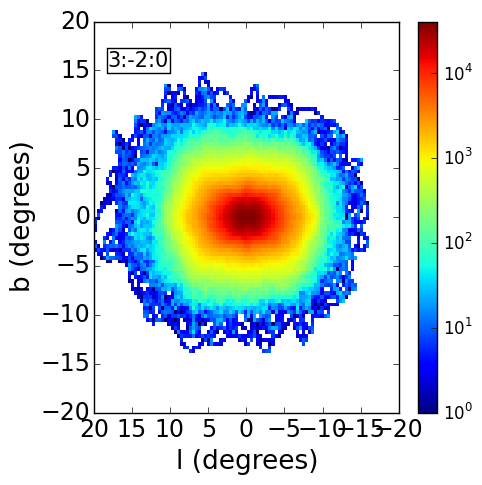}
	\includegraphics[width=.2\textwidth]{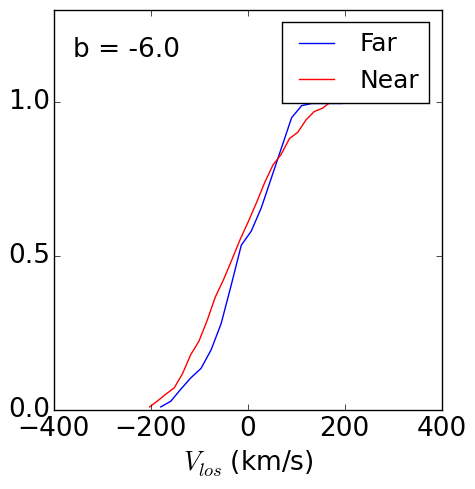}
	\includegraphics[width=.23\textwidth]{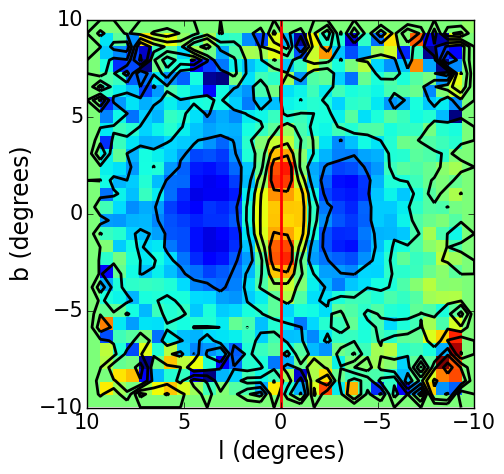}\\
%	\hspace{2cm}
	\includegraphics[width=.21\textwidth]{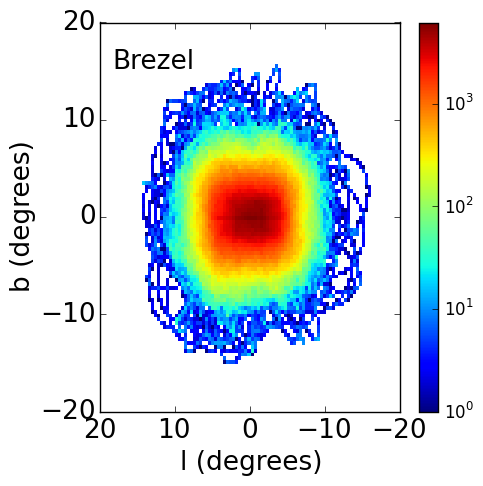}
	\includegraphics[width=.2\textwidth]{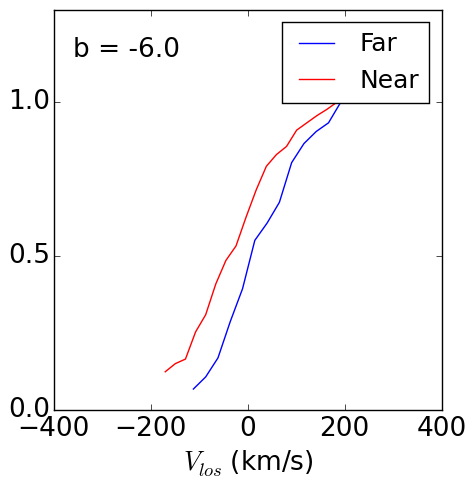}
	\includegraphics[width=.23\textwidth]{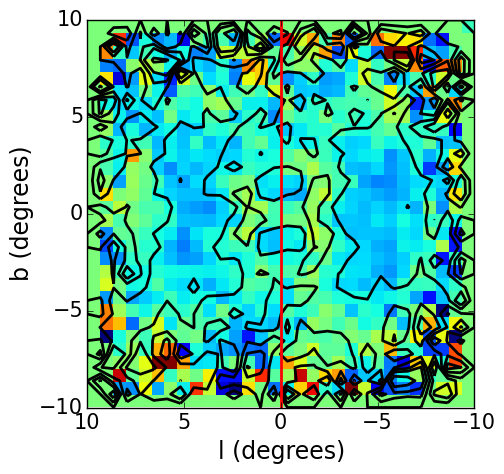}\\
\end{flushleft}
%\end{minipage}	
%\begin{minipage}
%\begin{center}
%\end{center}
%\end{minipage}	
    \caption{Left 3 columns: orbit projections as a function of $l, b$ for co-added orbits belonging to five families in Model A.  From top to bottom:  boxes, $x$-tubes, x1+banana orbits, 3:-2:0 (fish/pretzels),and 3:0:-5 (brezels).  The first column contains the projected density of the orbits viewed in the HCR frame.  The second column shows the CDF of \vlos for near (red) and far (blue) stars at $b=-6^{\circ}$.  The third column shows maps of the difference in the mean values of \vlos\ for all $l,b$ values in the bar region.  Lastly the image in the fourth column is the same as found in Fig. \ref{fig:Kin_A} in the fourth column.}
    \label{fig:no_orb}
\end{figure*}
%%%% Figure 8 -- Histograms, CDF and 2D velocity distribution maps %%%%%

As mentioned previously, \citet{vasquez_etal_13} showed, using line-of-sight velocities for RC stars in the MW bulge, that stars on the near side of the GC have more negative mean velocities than stars on the far side, a feature that is reproduced by Model A  (Fig.~\ref{fig:Kin_A}). We now compare these results with co-added orbits from  the five different families that were examined in previous sections. 

Each row of Figure ~\ref{fig:no_orb}  shows one co-added orbit family in the HCR frame. The first three columns show (from left to right), the projected density of stars, the CDFs of \vlos\ for stars on the `Near'  (red) and `Far' (blue) sides at $l=0^\circ, b = -6^\circ$, and  2D maps showing $\Delta V_{\rm los}$  over the same range of $l$ and $b$ as in Figure~\ref{fig:Kin_A}. (The 2D kinematic map for Model A is included in the 3rd row, 4th column to facilitate comparison with individual orbit families.) The first column of this figure shows that all the orbit families show some degree of asymmetry due to the projection effects arising from the orientation of the bar to our line-of-sight to the GC, but the x1+banana orbits show the greatest asymmetry (because they extend furthest out along the major axis).

The CDFs for individual orbits at $l=0^\circ, b = -6^\circ$  (2nd column of Fig. ~\ref{fig:no_orb}) show only a small difference between the `Near' and `Far' side  \vlos\, providing little ability to discriminate between the different orbit families. Most orbit families qualitatively resemble the CDFs of Model A  shown in Fig. ~\ref{fig:Kin_A}, with the `Near' (red) curve always leading the `Far' (blue). The only exception is the x1+banana orbits where the blue curve leads at negative velocities, crossing over  the red curve at \vlos$\sim 0$.  However the velocity difference between `Near' and `Far'  for all orbit families is  smaller than it is for the full models and in the Milky Way. It is not necessarily surprising that no individual orbit family shows the velocity difference observed in the Milky Way or the full simulations, since the bars in these systems are not comprised of single orbit families, but the similarity between the CDFs for different families makes this a poor diagnostic. Fortunately, column 3 of Figure~\ref{fig:no_orb} shows that there is much greater contrast between individual orbit families in 2D maps of $\Delta V_{\rm los}$ than there is at a specific position (e.g.  $l=0^{\circ}, b=-6^{\circ}$ in column 2).

 In fact \citet{gardner_etal_14} and \citet{qin_etal_15} showed that 2D maps of $\Delta V_{\rm los}$  for simulated bars with a strong BP/X-shape show an asymmetric  ring like structure with negative values of $\Delta V_{\rm los}$  surrounding a central region with slightly positive or zero values at $l=0^\circ, b=0^\circ$. In contrast  bar simulations without an X-shape show vertical contours of $\Delta V_{\rm los}$. 
 
A comparison with the  $\Delta V_{\rm los}$ map for  Model A (4th column, 3rd row in Fig.~\ref{fig:no_orb}) shows that no single orbit family fully accounts for the ring like structure in the 2D map for Model A. Boxes (top row), 3:-2:0 fish/pretzels (4th row) and 3:0:-5 brezels (5th row) show similar structures to the full model.  The $x$-tube orbits shows a velocity signature that has the opposite sign of velocity relative to other orbit families and relative to  Model A.  The x1+banana orbits (4th row) show a completely different 2D kinematic map  with much more strongly negative velocities in a butterfly shape. However only this last family produces the highly negative velocities at $5^\circ < l < 10^\circ$ and all $b$ values. As mentioned previously, this family is found predominantly in the outer half of the bar and at this range of $l$ values the near-end of the bar is quite close to the Sun, making this family very prominent in this part of the map.  

Once again we find, from a comparison of 2D maps of $\Delta V_{\rm los}$ of individual orbit families with the full simulation of Model A, that no single orbit family completely accounts for the BP/X-shape. The boxes, and 3:-2:0 and 3:0:-5 resonant boxlets have similar overall shape in the 2D kinematic maps and account for most of the structure seen in the simulation. While the x1+banana family has a distribution on this map that is completely different from that of the full model and the other orbits, only this family produces the high negative $\Delta V_{\rm los}$ values  seen in the 2D map at large $l$. Thus the 2D maps of the kinematics provide additional confirmation that all the non-resonant box and resonant boxlet families together contribute to the BP/X shape seen in the MW bulge.
		
\section{Summary and Conclusions}
\label{sec:concl}

In a previous paper (V16) we developed a new automated orbit classification algorithm to classify orbits in N-body bars using their fundamental orbital frequencies. In this work we have used our previously published orbit classification for an N-body bar (scaled to fit BRAVA kinematics of the Milky Way bulge), to determine which orbit family or families are primarily responsible for the box/peanut  and X-shape seen in the Milky Way bulge. Although the bar model we present in this paper was not tailored to fit observations in the Milky Way, it shows numerous similarities both in kinematics and in spatial distribution of stars contributing to the X-shape, enabling us to draw inferences about the nature of the orbits that might constitute the Milky Way bar. The results presented here should be followed up in the future by a similar analysis of a self-consistent model that is specifically tailored to fit the Milky Way bar. We assess the importance of various orbit families to the structure of the X-shape by examining four different diagnostics for the full simulation and the co-added orbit families: (1) the projected density distributions and unsharp masked images, (2) projected 3D distribution of the ``excess-mass'' outside ellipsoids , (3) the radial distribution of stars along  several lines-of-sight, and (4) 2D  maps showing $\Delta V_{\rm los}$, the difference in line-of-sight velocity between stars on the `Near' and `Far' sides of the Galactic Center. We summarize our main findings below.

\begin{enumerate}
\item{Our examination of co-added orbit families using edge-on projected density distributions and unsharp masked images (Fig.~\ref{fig:x_orb}) shows that the main contributors to the box/peanut shape in projected density maps  {for our N-body model} are non-resonant box orbits (which constitute $\sim$60\% of  bar orbits),  long axis ($x$) tubes (8.5\% of bar orbits),  and resonant boxlet families:  3:-2:0 ``fish/pretzel'' resonance (6\% of bar orbits), 3:0:-5 ``brezel'' orbits (1.5\%) and   2:-2:1 x1+banana orbits (3\%).}

\item{In contrast a clear X-shape  in unsharp masked images is only produced by the resonant boxlet families associated with the 3:-2:0  resonance  and the 3:0:-5 resonance.  While the resonant x1+banana  orbits show a clear box/peanut/X-shape in projection they show a double shell-like structures at $\pm 2$~kpc in unsharp masked images but no X-shape.}

\item{In a manner similar to that employed by \citet{portail_etal_15a} the fraction of mass associated with the X-shape is determined by computing several isodensity surfaces and subtracting the mass of an ellipsoid that lies entirely within each surface. This method yields a mass fraction associated with the BP/X-shape of 23\% in Model A and 19\% in Model C comparable to that found by \citep[][$24^{+5}_{-4}$\%]{portail_etal_15a}. Figure~\ref{fig:model_Xmass} shows that both models show a clear BP/X shape when the excess mass is projected onto the $x-z$ plane. A similar analysis carried out for co-added orbits in Model A finds excess masses ranging from 10\% for $x$-tubes  to 25\% for non-resonant boxes and 33\% for x1+banana orbits, confirming that all the orbit families considered contribute to the 3-dimensional BP/X-shape. Figure~\ref{fig:orbits_Xmass} shows that box orbits, x1+banana orbits, fish/pretzel orbits and brezels all show X-shapes in their 3D excess mass distributions, with each orbit family contributing to the X-shape over a different radial range.  }

\item{The bifurcation in the distributions in red clump star number counts as a function of observational magnitude \citep{mcwilliam_zoccali_10, nataf_10} for $|b| \gtrsim 5^\circ$ is a prominent signature of the 3-dimensional structure of the Milky Way's X-shaped bulge. In the MW bulge and the full simulation (top row of Figure~\ref{fig:radhist} and Table~\ref{tab:rhelio_sep}) the increasing separation of the peaks of the number count distribution with increasing $|b|$ is evidence that our line-of-sight through the MW bulge is passing through two opposing arms of the X-shape that get farther apart as  $|b|$ increases. A comparison with co-added orbit families shows that only the box orbit family shows such an increasing separation of peaks with increasing $|b|$ (Fig. \ref{fig:radhist} and Tab. \ref{tab:rhelio_sep}), while the resonant orbit families (3:-2:0 fish/pretzel, 3:0:-5 brezels, and 2:-2:1 x1+banana orbits)  show nearly constant separation in the peaks at the four $b$ values examined. Furthermore x1+banana orbits  show a much greater separation of peaks ($\sim 3$~kpc) at all $b$ values than observed in the full simulation. }

\item{The difference between the line-of-sight velocities of stars on the `Near' and `Far' sides of the GC in the Milky Way at $l=0^\circ, b=-6^\circ$ is observed to be about 50~\kms. We see a similar velocity signature but with a smaller velocity difference of $\sim 24$~\kms in the full simulations. Most of the co-added orbit families show a small velocity difference at this specific $l,b$ value (Fig. \ref{fig:no_orb} middle column), providing no discriminatory potential.}

\item{In 2D kinematic maps (Fig. \ref{fig:no_orb} right column) the velocity difference between `Near' and `Far' side stars for N-body simulations with X-shapes show a distinctive asymmetric ring shaped region of negative $\Delta V_{\rm los}$ surrounding a region with slightly positive or zero values at $l=0^\circ, b=0^\circ$ \citep{gardner_etal_14}. The 2D kinematic maps for the co-added orbits  for box, and boxlet families show overall similarity to  the structure seen in the full simulation except at $5^\circ <l <10^\circ$ (the near end of the bar) where only x1+banana orbits produce sufficiently negative $\Delta V_{\rm los}$.} 

\end{enumerate}

In summary our use of four different diagnostics to compare co-added orbit families with the full simulations show that the BP/X shape  {in our N-body model} is produced largely by the box orbit family which constitutes the majority of orbits in the bar. Although resonant families such as the 2:-2:-1 x1v1 (banana), 3:-2:0 fish/pretzels, 3:0:-5 brezels contribute to the box/peanut and X-shapes in projected density maps none of these families is individually able to account for the increasing separation with increasing $|b|$ of the radial stellar number counts and the asymmetric ring structure of the 2D kinematic maps. These results  {suggest} that the non-resonant box orbits, in conjunction with the resonant boxlets are {\it collectively} responsible for all the observed features of the BP/X-shape seen in the MW bulge.

It is not entirely surprising that the box orbits and resonant boxlets constitute the major families making up the box/peanut shape and the X-shapes seen in our bar simulation. Non-resonant box orbits are comprised of 3 independent fundamental orbital frequencies. \citet{valluri_etal_10}  studied  a triaxial potential primarily comprised of box orbits  as it was deformed by the addition of baryons, to  a more spherical/oblate potential. They found that despite the dramatic increase in oblateness at small radii, the overall orbit populations in this region did not change very significantly, but rather the box orbits  at small radii adiabatically deformed to become much rounder. In other words, since box orbits have three independent orbital fundamental frequencies, each frequency can be changed independently of the others, allowing these orbits to easily change shape as the underlying potential  is modified.

It is as yet uncertain whether the BP/X shape seen in edge-on bars arises from  the rapid buckling instability of the bar or whether it arises from more the adiabatically varying potential change associated with resonant trapping. Regardless of the process by which these structures form, since non-resonant box orbits  are the backbones of bars \citep{valluri_etal_16} they readily adapt to potential change in a manner similar to Lissajous figures. 

Recent analysis of the kinematics of $\sim2000$ giant stars in the direction of the Galactic bulge, obtained by the Gaia-ESO survey show differences in the line-of-sight velocity dispersions of metal poor and metal rich stars, which these authors argue point to the metal rich stars being on banana orbits \citep{williams_etal_16_metalrichbanana}. %\citet{debattista_etal_16_fractionation} present a novel mechanism for producing kinematic differences between populations of different metallicities. 
Our comprehensive analysis of orbits in an N-body bar shows that the banana orbit family is not capable of reproducing all spatial and kinematic characteristics associated with the X-shape. Future analysis of the orbits of stars in hydrodynamical simulations  of bars could lead to additional insights into correlations between metallicity, kinematics and orbit type, perhaps enabling us to distinguish between the two main formation mechanisms, bar buckling and resonant orbit trapping.

\section*{Acknowledgements}
 MV was supported in part by University of Michigan's Office of Research, HST-AR-13890.001, NSF awards AST-0908346, AST-1515001, NASA-ATP award NNX15AK79G.  CA  was supported by NSF-AST-1515001. CA and MV also thank Sarah Loebman and Eric Bell for numerous helpful discussions during the course of this work.  J.S. is partially supported by the 973 Program of China under grant no. 2014CB845700, by the National Natural Science Foundation of China under grant nos.11333003, 11322326, and by a China-Chile joint grant from CASSACA. J.S. also acknowledges support from an {\it Newton Advanced Fellowship} awarded by the Royal Society and the Newton Fund, and from the CAS/SAFEA International Partnership Program for Creative Research Teams. This work made use of the facilities of the Center for High Performance Computing at Shanghai Astronomical Observatory. VPD is supported by STFC Consolidated grant \#~ST/M000877/1 and partially supported by the Chinese Academy of Sciences President's International Fellowship Initiative (PIFI, Grant No. 2015VMB004).

%%%%%%%%%%%%%%%%%%%%%%%%%%%%%%%%%%%%%%%%%%%%%%%%%%

%%%%%%%%%% 2d Projection and unsharp mask



\bibliographystyle{mnras}
\bibliography{Abbott_xshape_bib} % if your bibtex file is called example.bib
%\include{Abbott_etal_xshapeV1.bbl}

%\end{thebibliography}

%%%%%%%%%%%%%%%%%%%%%%%%%%%%%%%%%%%%%%%%%%%%%%%%%%

%%%%%%%%%%%%%%%%% APPENDICES %%%%%%%%%%%%%%%%%%%%%

%%%%%%%%%%%%%%%%%%%%%%%%%%%%%%%%%%%%%%%%%%%%%%%%%%
% Don't change these lines
\bsp	% typesetting comment
\label{lastpage}
\end{document}